\documentclass[11pt]{article}

\usepackage{amssymb} 
\usepackage{amsmath}     
\usepackage{amsthm}
\usepackage{todonotes}
\usepackage{mathtools}
\usepackage{authblk}

\usepackage[ruled,vlined,linesnumbered]{algorithm2e}
\SetKw{KwNot}{not}
\SetKw{KwOR}{or}
\SetKw{KwAND}{and}
\SetKw{KwExitFor}{exit for-loop}

\usepackage{a4wide}
\usepackage{graphicx}
\usepackage{hyperref}

\newcommand{\rvY}{{\rm Y}^{\pmb{x}}}

\newtheorem{thm}{Theorem}
\newtheorem{lem}{Lemma}

\newtheorem{cor}{Corollary}

\newtheorem{obs}{Observation}

\begin{document}

\title{Two-stage Combinatorial Optimization Problems under Risk}

\author[1]{Marc Goerigk}
\author[2]{Adam Kasperski}
\author[3]{Pawe{\l} Zieli\'nski}

\affil[1]{Network and Data Science Management, University of Siegen, Germany\\
           \texttt{marc.goerigk@uni-siegen.de}}
\affil[2]{Faculty of Computer Science and Management, Wroc{\l}aw University of Technology, Poland\\
            \texttt{adam.kasperski@pwr.edu.pl}}
\affil[3]{Faculty of Fundamental Problems of Technology, Wroc{\l}aw University of Technology, Poland\\
       \texttt{pawel.zielinski@pwr.edu.pl}}
   
    \date{}
    
\maketitle

 \begin{abstract}
 In this paper a class of combinatorial optimization problems is discussed. It is assumed that a solution can be constructed in two stages. The current first-stage costs are precisely known, while the future second-stage costs are only known to belong to an uncertainty set, which contains a finite number of scenarios with known probability distribution. A partial solution, chosen in the first stage, can be completed by performing an optimal recourse action, after the true second-stage scenario is revealed. A solution minimizing the Conditional Value at Risk (CVaR) measure is computed. Since expectation and maximum are boundary cases of CVaR, the model generalizes the traditional stochastic and robust two-stage approaches, previously discussed in the existing literature. In this paper some new negative and positive results are provided for basic combinatorial optimization problems such as the selection or network problems.
 
\end{abstract}

\textbf{Keywords:} combinatorial optimization; stochastic programming; two-stage problems; optimization under risk; robust optimization


\section{Introduction}

In many applications of optimization models a solution can be constructed in two stages. Namely, a partial solution is formed in the first stage and completed in the second stage. 
 Typically, the current first stage costs are precisely known, while the future second stage costs are uncertain. We can only predict that the second stage costs will belong to an \emph{uncertainty (scenario) set} $\mathcal{U}$. In a corresponding mathematical programming model we have two sets of variables. The first set contains the variables which are fixed \emph{now}, i.e before a true second-stage scenario reveals. The second set contains variables which can be fixed \emph{in the future}, i.e. when the true scenario from $\mathcal{U}$ becomes known. When a probability distribution in $\mathcal{U}$ is available, then a stochastic approach can be applied to solve the problem. Namely, we seek a solution minimizing a given risk measure (in most papers the expectation is used). The two-stage models have a long tradition in stochastic programming, in particular in stochastic linear programming~(see, e.g~\cite{KM05, KW94}). In practice, a probability distribution in $\mathcal{U}$ can be unknown. In this case we can use the \emph{robust optimization} framework~(see, e.g.~\cite{BN09}), in which we assume that a worst scenario from $\mathcal{U}$ will occur in the second stage. The two-stage approach is then similar to the concept of \emph{adjustable robustness} in mathematical programming~\cite{BTG03}, where part of the variables must be determined before the realization of the uncertain parameters, while the other part are variables that can be chosen after the realization.

In combinatorial optimization problems we typically seek an object composed of elements of some finite set $E$. For example, we are given a graph $G=(V,E)$ and we wish to find a cheapest subset of the edges forming an $s-t$ path, spanning tree, assignment etc. in $G$. In this case, the two-stage approach emerges naturally. A subset of the elements, forming a partial solution is chosen in the first stage, and in the second stage this subset is completed to a feasible solution. Particular two-stage combinatorial optimization problems, namely the two-stage spanning tree, assignment and selection problems, have been recently discussed in~\cite{KMU08, FFK06, DRM05, KZ11, KZ15b}. The problems were investigated in both stochastic and robust settings. In the models discussed, scenario set $\mathcal{U}$ contains a finite number of~$K$ scenarios. In the stochastic models considered in~\cite{KMU08, FFK06, DRM05}, the expectation was used as a risk measure. It is well known that the expectation is sometimes not appropriate criterion, as it does not take decision makers risk-aversion into account. So, a more sophisticated risk measure should be used. On the other hand, the robust min-max criterion is often regarded as too conservative.

In this paper we investigate two-stage versions of some basic combinatorial optimization problems. We assume that scenario set $\mathcal{U}$ contains $K$ scenarios with known probability distribution. A partial solution, computed in the first stage, can be completed in the second stage by performing an optimal \emph{recourse action}, after the true scenario is revealed. 
The goal is to find a first stage solution minimizing a certain risk measure. As a risk measure, we will use the \emph{Conditional Value at Risk} (CVaR for short)~\cite{P00,RU00}, which is a popular criterion in stochastic optimization, which takes decision maker risk-aversion into account. The expectation and the maximum can be seen as boundary cases of CVaR, so using this measure allows us to establish a link between the traditional stochastic and robust approaches and generalize the problems investigated in~\cite{KMU08, DRM05, KZ11}. 

This paper is organized as follows. We introduce the two-stage combinatorial optimization setting in Section~\ref{secproblem}. In Section~\ref{secgenprop} we show some general properties, which describe the complexity of minimizing CVaR, if some results for the expectation and maximum criteria are available. In Sections~\ref{sechs} and~\ref{secsel}, we consider two very basic selection problems, which can be solved trivially in the deterministic case. We show that in the two-stage approach both selection problems can be hard to solve and also hard to approximate. We then investigate the class of basic network problems, namely the shortest path, minimum spanning tree and minimum assignment in Section~\ref{secnetworks}. We strengthen the known results in this area and provide new ones. In particular, we consider several variants of the two-stage shortest path problem, which have various complexity properties. In order to characterize the complexity of the network problems in some very restrictive cases, we use the results obtained for the selection problems. The table summarizing the obtained results and containing some open problem is presented in the concluding Section~\ref{secconclusions}.

\section{Problem formulation}
\label{secproblem}

In this paper, we consider the following combinatorial optimization problem $\mathcal{P}$:
$$\begin{array}{llll}
		\min & \pmb{C}\pmb{x} \\
			& \pmb{x}\in \mathcal{X} \subseteq \{0,1\}^n,\\
	\end{array}
	$$
where $\pmb{C}=[C_1,\dots,C_n]$ is a vector of nonnegative costs and $\mathcal{X}$ is a set of feasible solutions, which is typically a set of characteristic vectors of some element set (tools, items, arcs of some network etc.). 
 Given a vector $\pmb{x}\in \{0,1\}^n$, let us define the following set of \emph{recourse actions}:
$$\mathcal{R}(\pmb{x})=\{\pmb{y}\in \{0,1\}^n: \pmb{x}+\pmb{y}\in \mathcal{X}\}$$
and a set of \emph{partial solutions} is defined as follows:
$$\mathcal{X}'=\{\pmb{x}\in \{0,1\}^n: \mathcal{R}(\pmb{x})\neq \emptyset\}.$$

Observe that $\mathcal{X}\subseteq \mathcal{X}'$ and $\mathcal{X}'$ contains all vectors which can be completed to a feasible solution in $\mathcal{X}$. Sometimes additional restrictions can be imposed on $\mathcal{X}'$. We will consider one such a case in Section~\ref{secsp}.
A partial solution $\pmb{x}\in \mathcal{X}'$ is  completed in the second stage, i.e. we choose $\pmb{y}\in \mathcal{R}(\pmb{x})$ which yields $\pmb{x}+\pmb{y}\in \mathcal{X}$. The overall cost of the solution constructed is $\pmb{C}\pmb{x}+\pmb{c}\pmb{y}$ for a fixed cost vector $\pmb{c}$. We will assume that the vector of the first-stage costs $\pmb{C}$ is known but the vector of the second-stage costs 
$\pmb{c}$
is uncertain and belongs to a specified discrete uncertainty (scenario) set $\mathcal{U}=\{\pmb{c}_1,\dots,\pmb{c}_K\}$, where scenario $\pmb{c}_j$ occurs with probability $p_j>0$. We will use $c_{ij}$ to denote the second-stage cost of the $i$th variable under scenario $j$.
Fix $\pmb{x}\in \mathcal{X}'$ and define
$$\pmb{y}_j^{\pmb{x}}={\rm arg} \min_{\pmb{y}\in \mathcal{R}(\pmb{x})} \pmb{c}_j \pmb{y},$$
i.e. $\pmb{y}_j^{\pmb{x}}$ is an \emph{optimal recourse action} for $\pmb{x}$ under $\pmb{c}_j$. Each vector $\pmb{x}\in \mathcal{X}'$ induces a discrete random variable $\rvY$ which takes the value of $\pmb{c}_j\pmb{y}_j^{\pmb{x}}$ with probability $p_j$. In this paper we investigate the following \emph{two-stage problem}:
$$\textsc{TSt-F}~\mathcal{P}: \min_{\pmb{x}\in \mathcal{X}'} (\pmb{C}\pmb{x}+{\bf F}[\rvY]),$$
where ${\bf F}$ is a given risk measure. Will use the following popular risk measure, called \emph{Conditional Value at Risk}~\cite{P00,RU00}:
$${\bf CVaR}_{\alpha}[\rvY]=\inf\{\gamma+\frac{1}{1-\alpha}{\bf E}[\rvY-\gamma]^+: \gamma\in R\},\; \alpha \in [0,1).$$
The problem $\textsc{TSt-CVaR}_{\alpha}~\mathcal{P}$ can be formulated as the following MIP model:
$$
	\begin{array}{lllll}
		\min & \displaystyle \pmb{C}\pmb{x}+\gamma + \frac{1}{1-\alpha}\sum_{j\in [K]} p_j u_j \\
			& u_j \geq \pmb{c}_j\pmb{y}_j-\gamma & j\in [K]\\
			& \pmb{x}+\pmb{y}_j\in \mathcal{X} & j\in [K] \\
			& u_j\geq 0 & j\in [K]
	\end{array}
$$
where we use the notation $[K]:=\{1,\ldots,K\}$.
In this formulation $\pmb{y}_j$ is an optimal recourse action associated with $\pmb{x}$ and scenario $\pmb{c}_j$. If $\alpha=0$, then we get the following \emph{stochastic two-stage problem}:
$$\textsc{TSt-E}~\mathcal{P}:\min_{\pmb{x}\in \mathcal{X}'} (\pmb{C}\pmb{x}+{\bf E}[\rvY]),$$
Where $\textbf{E}$ denotes the expectation.
On the other hand, when $\alpha\rightarrow 1$, we get the following \emph{robust two-stage problem}:
$$\textsc{TSt-R}~\mathcal{P}:\min_{\pmb{x}\in \mathcal{X}'} (\pmb{C}\pmb{x}+\max_{j\in [K]} \pmb{c}_j\pmb{y}_j^{\pmb{x}}).$$
Notice that in the robust problem, the scenario probabilities are ignored.

\begin{figure}[ht]
	\centering
	\includegraphics[height=2.5cm]{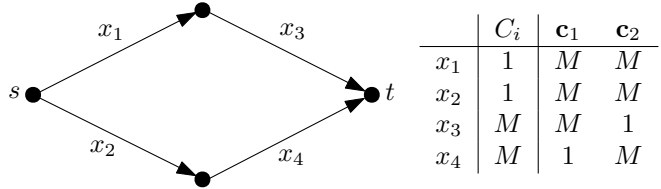}
	\caption{An instance of  \textsc{TSt-R Shortest Path} with two scenarios, $M$ is a big constant.} \label{fig3} 
	\end{figure}

Let $\overline{\mathcal{X}}=\{\pmb{x}+\pmb{z} \in\{0,1\}^n: \pmb{x}\in \mathcal{X}, \pmb{z}\in \{0,1\}^n\}$, hence $\overline{\mathcal{X}}$ contains all vectors $\pmb{x}\in \mathcal{X}$ and all vectors obtained by replacing one or more~0 components with~1 in $\pmb{x}\in \mathcal{X}$. If all the costs are nonnegative, then the deterministic, single-stage problems $\mathcal{P}$ with $\mathcal{X}$ and $\overline{\mathcal{X}}$ are equivalent. However, this is not the case in the two-stage approach when we use $\overline{\mathcal{R}}(\pmb{x}) = \left\{ \pmb{y}\in\{0,1\}^n : \pmb{x}+\pmb{y}\in\overline{\mathcal{X}}\right\}$. To see this consider an instance of the \textsc{TSt-R Shortest Path} problem shown in Figure~\ref{fig3}. If $\mathcal{X}=\{(1,0,1,0), (0,1,0,1)\}$, i.e. it contains only characteristic vectors of the two $s-t$ paths, then the cost of the optimal solution is $M+1$. However, if we extend the set of feasible solutions to $\overline{\mathcal{X}}$, then the cost of an optimal solution becomes~3. So, it can be advantageous to select some redundant arcs in the first stage, which will be not used in the second stage. Contrary to the deterministic case, the problems with $\mathcal{X}$ and $\overline{\mathcal{X}}$ can have different computational properties in the two-stage setting.

\section{Some general properties of the problem}
\label{secgenprop}

We note first that $\textsc{TSt-CVaR}_{\alpha}~\mathcal{P}$ has the same complexity as $\mathcal{P}$ if $K=1$, i.e. when the second stage costs are known precisely and are equal to $\pmb{c}$ (scenario $\pmb{c}$ occurs with probability equal to~1).  Let $\hat{c}_i=\min\{C_i, c_i\}$ for $i\in [n]$ and let $\hat{\pmb{x}}\in \mathcal{X}$ be an optimal solution to~$\mathcal{P}$ for the costs $\hat{c}_i$, $i\in [n]$. We then fix $x_i=1$ if $\hat{x}_i=1$ and $\hat{c}_i=C_i$. This gives us a feasible solution $\pmb{x}\in \mathcal{X}'$, which can be completed under scenario $\pmb{c}$ by fixing $y_i=1$ if $\hat{x}_i=1$ and $\hat{c}_i=c_i\neq C_i$, $i\in [n]$. It is clear that $\pmb{x}$ is an optimal first-stage solution and $\pmb{y}\in \mathcal{R}(\pmb{x})$ is an optimal recourse action for $\pmb{x}$. 

\begin{lem}[\cite{KZ19a}]
\label{lem01}
	Let $\mathrm{Y}$ be a discrete random variable which takes $K$ nonnegative values
	with probabilities $p_1,\dots, p_K$.
	The following inequalities hold for each $\alpha\in [0,1)$:
	\begin{equation}
		{\bf E}[\mathrm{Y}]\leq {\bf CVaR}_\alpha [\mathrm{Y}] \leq
		 \min\left \{\frac{1}{{\Pr}_{\min}},\frac{1}{1-\alpha} \right\} {\bf E}[\mathrm{Y}],
	\end{equation}
	where ${\rm Pr}_{\min}=\min_{j\in [K]} p_j$.
\end{lem}

\begin{thm}
\label{thmgen0}
	If \textsc{TSt-E}~$\mathcal{P}$ is approximable within $\rho>1$ (for $\rho=1$ it is polynomially solvable), then $\textsc{TSt-CVaR}_{\alpha}~\mathcal{P}$ is approximable within $\rho\sigma$, where
	$\sigma= \min\left \{\frac{1}{{\Pr}_{\min}},\frac{1}{1-\alpha} \right\}$ for $\alpha\in [0,1)$.
\end{thm}
\begin{proof}
	Let $\hat{\pmb{x}}$ and $\pmb{x}^*$  be optimal solutions to \textsc{TSt-E}~$\mathcal{P}$ and $\textsc{TSt-CVaR}_{\alpha}~\mathcal{P}$, respectively. Let $\overline{\pmb{x}}$ be the solution produced by the $\rho$-approximation algorithm for \textsc{TSt-E}~$\mathcal{P}$.
	Using Lemma~\ref{lem01}, we get
{\begin{align*}
\pmb{C}\overline{\pmb{x}}+{\bf CVaR}_{\alpha}[{\rm Y}^{\overline{\pmb{x}}}]
&\leq \pmb{C}\overline{\pmb{x}}+\sigma {\bf E}[{\rm Y}^{\overline{\pmb{x}}}] 
\leq \rho\sigma(\pmb{C}\hat{\pmb{x}}+ {\bf E}[{\rm Y}^{\hat{\pmb{x}}}])\\
&\leq \rho\sigma(\pmb{C}\pmb{x}^*+ {\bf E}[{\rm Y}^{\pmb{x}^*}])
\leq \rho\sigma(\pmb{C}\pmb{x}^*+ {\bf CVaR}_{\alpha}[{\rm Y}^{\pmb{x}^*}])
\end{align*}}
	and the theorem follows. 
\end{proof}

\begin{lem}[\cite{KZ19a}]
\label{lem02}
	Let $\mathrm{Y}$ be a discrete random variable which takes $K$ nonnegative values $b_1,\dots, b_K$ with probabilities $p_1,\dots, p_K$.  Let us create random variable ${\rm Y}'$ which takes values $0, b_1,\dots,b_K$ with probabilities $\alpha, p_1(1-\alpha),\dots,p_K(1-\alpha)$ for some  fixed $\alpha\in [0,1)$. Then
	$${\bf E}[{\rm Y}]={\bf CVaR}_{\alpha}[{\rm Y}'].$$
\end{lem}

\begin{thm}
\label{thmehard}
The following statements hold:
	\begin{itemize}
	\item  If $\textsc{TSt-E}~\mathcal{P}$ with $K$ scenarios is (strongly) NP-hard, then  problem $\textsc{TSt-CVaR}_{\alpha}~\mathcal{P}$ with $K+1$ scenarios is also  (strongly) NP-hard for any fixed $\alpha \in [0,1)$.
	\item If $\textsc{TSt-E}~\mathcal{P}$ with $K$ scenarios is hard to approximate within $\rho$, then  problem $\textsc{TSt-CVaR}_{\alpha}~\mathcal{P}$ with $K+1$ scenarios is also hard to approximate within $\rho$, for any fixed $\alpha \in [0,1)$.
	\end{itemize}
\end{thm}
\begin{proof}
	Consider an instance $I=(\mathcal{X},\pmb{C},\mathcal{U})$ of $\textsc{TSt-E}~\mathcal{P}$. Fix $\alpha\in [0,1)$ and define the instance $I'=(\mathcal{X},\pmb{C},\mathcal{U}')$ of $\textsc{TSt-CVaR}_{\alpha}~\mathcal{P}$ with $\mathcal{U}'= \mathcal{U} \cup \{\pmb{c}_0\}$, where $\pmb{c}_0$ is a vector of zeros. The probability of $\pmb{c}_0$ is $\alpha$ and the probability of $\pmb{c}_j$ is $p_j(1-\alpha)$, $j\in [K]$, in $I'$. Both instances have the same set of feasible solutions $\mathcal{X}'$. A solution $\pmb{x}\in \mathcal{X}'$ induces random variable $\rvY$ in $I$, taking the values $\pmb{c}_1\pmb{y}_1^{\pmb{x}},\dots, \pmb{c}_K\pmb{y}_K^{\pmb{x}}$ with probabilities $p_1,\dots,p_K$ and random variable ${\rm Y}'^{\pmb{x}}$ in $I'$ taking the values $0,\pmb{c}_1\pmb{y}_1^{\pmb{x}},\dots, \pmb{c}_K\pmb{y}_K^{\pmb{x}}$ with probabilities $\alpha, (1-\alpha)p_1,\dots,(1-\alpha)p_K$. From Lemma~\ref{lem02}, we get $\pmb{C}\pmb{x}+{\bf E}[\rvY]=\pmb{C}\pmb{x}+{\bf CVaR}_{\alpha}[{\rm Y}'^{\pmb{x}}]$. So, $I$ and $I'$ have optimal solutions with the same costs. In consequence, there is a cost preserving reduction from $\textsc{TSt-E}~\mathcal{P}$ with $K$ scenarios to $\textsc{TSt-CVaR}_{\alpha}~\mathcal{P}$ with $K+1$ scenarios and the theorem follows.
\end{proof}


\section{Complexity of two-stage representatives selection problem}
\label{sechs}

In this section we consider the following problem~$\mathcal{P}$, called \textsc{Representatives selection} (\textsc{RS} for short). We are given $\ell$ disjoint sets of tools $T_1,\dots, T_{\ell}$, which form a partition of the set $[n]=\{1,\dots,n\}$,  i.e. $T_i\cap T_j=\emptyset$ if $i\neq j$ and $T_1\cup \dots \cup T_{\ell}=[n]$. We wish to chose exactly one tool from each set to minimize the total cost of the selected tools.  So, the set of feasible solutions is defined as  $\mathcal{X}=\{\pmb{x}\in \{0,1\}^n: \sum_{i\in T_j} x_i=1, j\in [\ell]\}$. The robust single-stage version of RS was discussed in~\cite{DK12, DW13, KKZ15}, while a recoverable robust version was discussed in~\cite{B11}.

We note first that any instance of $\textsc{TSt-F RS}$ can be transformed in polynomial time to an equivalent instance of this problem in which $|T_l|=1$ for each $l\in [\ell]$. Given an instance $(\pmb{C}, \mathcal{U}, (T_l)_{l\in [\ell]})$, we only have to decide whether to choose an item in $T_l$ in the first or in the second stage. Notice that this choice is independent for each $l\in [\ell]$.  In the first case, we choose an item $i_l\in T_l$ of the smallest first stage cost and in the second case we choose the cheapest item from $T_l$ under each scenario. In consequence, we can create an equivalent instance of the problem $(\pmb{C}',\mathcal{U}', (T_l')_{l\in [\ell]})$, in which $T'_l$ contains only item $i_l\in T_l$, whose first stage cost equals $\min_{i\in T_l} C_i$ and $c_{i_l j}'=\min_{i\in T_l} c_{ij}$, $j\in [K]$. The number of scenarios in $\mathcal{U'}$ and the probability distribution is the same as in $\mathcal{U}$.  Hence, from now on, we will consider the case in which $|T_l|=1$ for each $l\in [\ell]$ and the problem is to decide whether tool $i\in [\ell]$ is chosen in the first or in the second stage.

\begin{obs}
\label{obs1}
\textsc{TSt-E RS} is polynomially solvable. 
 \end{obs}
 \begin{proof}
Consider the objective function of the \textsc{TSt-E RS} problem with $|T_l|=1$ for each $l\in [\ell]$:
\[ \pmb{C}\pmb{x} + {\bf E}[\rvY] =
\pmb{C}\pmb{x}+\sum_{j\in [K]} p_j \sum_{i\in [n]} c_{ij} (1-x_i) =\pmb{C}\pmb{x}+\sum_{i\in [n]} (1-x_i)\sum_{j\in [K]} p_j c_{ij}=\sum_{i\in [n]} (C_i x_i + \hat{c}_i (1-x_i)),\]
where  $\hat{c}_i =\sum_{j\in [K]} p_j c_{ij}$.
 We can now set $x_i=1$ if $C_i\leq \hat{c}_i$ and $x_i=0$, otherwise. This gives us an optimal solution to \textsc{TSt-E RS}, which can be easily constructed in polynomial time.
\end{proof}

The next results show that the robust version of the problem is much harder.

\begin{thm}
\label{thmrs1}
	\textsc{TSt-R RS} is NP-hard, even if $K=2$.
\end{thm}
\begin{proof}
	It is a direct consequence of  the result proven in~\cite[Theorem~5]{KZ15b}.
\end{proof}

\begin{cor}
	$\textsc{TSt-CVaR}_{\alpha}~\textsc{RS}$ is NP-hard for each $\alpha \in [0.5,1)$, even if $K=2$.
	\label{cors1}
\end{cor}
\begin{proof}
	Consider an instance of \textsc{TSt-R RS} with $K=2$ scenarios, which is NP-hard (see Theorem~\ref{thmrs1}). This problem is equivalent to $\textsc{TSt-CVaR}_{\alpha}~\textsc{RS}$ for any fixed $\alpha\in [0.5,1)$, defined for the same instance with probabilities $0.5$ for both scenarios.
\end{proof}

The complexity of the problem for $\alpha \in (0, 0.5)$ is open.

\begin{thm}
\label{thmrs2}
	\textsc{TSt-R RS} is strongly NP-hard for unbounded $K$.
\end{thm}
\begin{proof}
	 Consider the following \textsc{Min-Set Cover} problem. We are given a finite set $U = \{u_1,\dots, u_n\}$ and a collection $\mathcal{A}=\{A_1,\dots,A_m\}$ of subsets of $U$. A subset $D\subseteq \mathcal{A}$ covers $U$ (it is called a \emph{cover}) if for each element $u_i\in U$, there exists $A_j\in D$ such that $u_i\in A_j$. We seek a cover $D$ of the smallest size $|D|$. \textsc{Min-Set Cover} is known to be strongly NP-hard (see, e.g,~\cite{GJ79}). Given an instance of \textsc{Min-Set Cover} we construct the corresponding instance of \textsc{TSt-R RS} as follows.  Fix $M=n+1$. For each set $A_i\in \mathcal{A}$ we create a tool $i$ with the first stage cost equal to $M$. We also add an additional tool $n+1$ with the first stage cost equal to $2M$. For each element $u_j\in U$ we create scenario $\pmb{c}_j$ such that $c_{ij}=0$ if set $A_i$ contains $u_j$ and $c_{ij}=M$ otherwise. The cost of the additional $(n+1)$th tool is equal to $2M$. Finally, we add scenario $\pmb{c}'$ under which the cost of each tool in $[n]$ equals $M+1$ and the cost of the $(n+1)th$ tool is equal to $M$ (a sample reduction is shown in Table~\ref{tab1}).
	 
	 \begin{table}[ht]
	  \centering
	  \small
	  \caption{The instance of \textsc{TSt-R RS} for $U=\{u_1,\dots,u_7\}, \mathcal{A}=\{\{u_2,u_4,u_3\}, \{u_1\}, \{u_3,u_7\}, \{u_1,u_4,u_6,u_7\}, \{u_2,u_5,u_6\}, \{u_1,u_6\}\}$} \label{tab1}
			\begin{tabular}{l|c|cccccccc}
									 & $C_i$ & $\pmb{c}_1$ & $\pmb{c}_2$ & $\pmb{c}_3$ &  $\pmb{c}_4$ & $\pmb{c}_5$ & $\pmb{c}_6$ & $\pmb{c}_7$ & $\pmb{c}'$ \\ \hline
					$1$ 				& $M$  & $M$ & $0$ & $0$ & $0$ & $M$ & $M$& $M$ & $M+1$ \\
					$2$  				& $M$ & $0$ & $M$ & $M$ & $M$ & $M$ & $M$ & $M$ & $M+1$ \\  
					$3$ 				& $M$ & $M$ & $M$ & $0$ & $M$ & $M$& $M$ & $0$ & $M+1$ \\
					$4$ 				& $M$ & $0$ & $M$ & $M$ & $0$ & $M$ & $0$& $0$ & $M+1$\\  
					$5$				& $M$ & $M$ & $0$ & $M$ & $M$ & $0$& $0$ & $M$ & $M+1$\\
					$6$ 				& $M$ & $0$ & $M$ & $M$ & $M$ & $M$ & $0$ & $M$ & $M+1$  \\  \hline
					$7$ 				& $2M$ & $2M$ & $2M$   & $2M$ & $2M$& $2M$& $2M$& $2M$ & $M$\\
			\end{tabular}
		\end{table}

	 We will show that there is a cover $D$ of size at most $L$ if and only if there is a solution to \textsc{TSt-R RS} with the maximum total first and second stage cost at most $(n+1)M+L$. Assume first that there is a cover $D$ such that $|D|\leq L$. In the first stage we choose the tools from $[n]$ which do not correspond to the cover $D$. Since the not selected tools form a cover, the total first and second stage cost under scenario $\pmb{c}_j$ is at most $(n+1)M$ and under scenario $\pmb{c}'$ it is equal to $(n+1)M+L$. Hence the maximum total first and second stage cost of the tool selection is equal to $(n+1)M+L$.
	 
	 Assume not that there is a solution to \textsc{TSt-R RS} with the maximum cost a most $(n+1)M+L$. By the construction of scenario $\pmb{c}'$, we can conclude that at most $L$ tools from $[n]$ can be selected in the second stage. These tools correspond to the set $D\subseteq A$. Suppose that $D$ is not a cover and an element $u_j$ is not covered by $D$. Then the total cost of the selection under $\pmb{c}_j$ is $(n+2)M>(n+1)M+L$ (note that $M>L$), a contradiction. In consequence, there is a cover of size at most $L$.
\end{proof}

\begin{thm}
	$\textsc{TSt-CVaR}_{\alpha}$ \textsc{RS} is approximable within~$\min\{2,\frac{1}{1-\alpha}\}$.
	\label{tcvarrs2}
\end{thm}
\begin{proof}
	We note first that the  problem can be represented as the following MIP formulation:
\begin{equation}
\label{miprs}
	\begin{array}{lllll}
		\min & \displaystyle \pmb{C}\pmb{x}+\gamma + \frac{1}{1-\alpha}\sum_{j\in [K]} p_j u_j \\
			& \displaystyle u_j \geq \sum_{i\in [n]} c_{ij} (1-x_i)-\gamma & j\in [K]\\
			& u_j\geq 0 & j\in [K] \\
			& x_i\in\{0,1\} & i\in [n]
	\end{array}
\end{equation}

	We first show that the problem is approximable within~2. Consider a relaxation of~(\ref{miprs}) in which $x_i\in \{0,1\}$ is replaced with $0\leq x_i\leq 1$ for each $i\in [n]$. Let $\pmb{x}^*$ be an optimal solution to the relaxation. Fix $\pmb{y}^*=\pmb{1}-\pmb{x}^*$.  Define random variable ${\rm Y}^{\pmb{x}^*}$ taking values $\pmb{c}_1\pmb{y}^*, \dots,\pmb{c}_K\pmb{y}^*$ with probabilities $p_1,\dots, p_K$. The objective value of the relaxation of~(\ref{miprs}) can be then expressed as
	$$LB=\pmb{C}\pmb{x}^*+{\bf CVaR}_{\alpha}[{\rm Y}^{\pmb{x}^*}].$$
The equalities $x_i^*+y_i^*=1$ imply $x^*_i\geq 0.5$ or $y^*_i\geq 0.5$ for each $i\in [n]$. We now form a feasible solution by 
setting  $\hat{x}_i=1$ if $x^*_i\geq 0.5$ and $\hat{y}_i=1$ otherwise, for each $i\in [n]$. Since the conditional value at risk is monotone and homogeneous, i.e.
for two discrete random variables   $\mathrm{X}$ and $\mathrm{Y}$ taking nonnegative values
		 $a_1,\dots, a_K$, and $b_1,\dots, b_K$, respectively, with 
		${\rm Pr}[\mathrm{X}=a_i]={\rm Pr}[\mathrm{Y}=b_i]$ and $a_i\leq \gamma b_i$ for each $i\in [K]$ and some fixed $\gamma \geq 0$, the inequality  ${\bf CVaR}_\alpha[\mathrm{X}]\leq \gamma {\bf CVaR}_\alpha[\mathrm{Y}]$ holds
(see, e.g.,~\cite{KZ19a}),
 we get:
	$$\pmb{C}\hat{\pmb{x}}+{\bf CVaR}_{\alpha}[{\rm Y}^{\hat{\pmb{x}}}]\leq 2\pmb{C}\pmb{x}^*+{\bf CVaR}_{\alpha}[2{\rm Y}^{\pmb{x}^*}]=2(\pmb{C}\pmb{x}^*+{\bf CVaR}_{\alpha}[{\rm Y}^{\pmb{x}^*}])= 2\cdot LB,$$
	so the rounded solution is a 2-approximate one. The approximation ratio $\frac{1}{1-\alpha}$ follows from Theorem~\ref{thmgen0} and the fact that \textsc{TSt-E RS} is polynomially solvable.
\end{proof}
\begin{thm}
	\textsc{TSt-R RS} is approximable within~2.
	\label{trrs2}
\end{thm}
\begin{proof}
The robust problem can be stated as the following MIP model:
\begin{equation}
\label{miprsr}
\begin{array}{lllll}
		\min & L \\
			& \pmb{C}\pmb{x}+\sum_{i\in [n]} c_{ij} (1-x_i)\leq L& j\in [K]\\
			& x_i\in\{0,1\} & i\in [n]
	\end{array}
\end{equation}
Analogously to the proof of Theorem~\ref{tcvarrs2}, we solve the LP relaxation of~(\ref{miprsr}) and round the optimal solution $(\pmb{x}^*,\pmb{y}^*)$ to an integer solution with at most twice the objective value of the LP solution.
\end{proof}


\section{Complexity of two-stage selection problem}
\label{secsel}

In this section we consider the case in which $\mathcal{P}$ is the following \textsc{Selection} problem. We are given as set of $[n]$ items and we wish to choose exactly $p$ out of them to minimize their total cost. Hence $\mathcal{X}=\{\pmb{x}\in \{0,1\}^n: x_1+\dots+x_n=p\}$, where $p\in [n]$. 
The \textsc{TSt-R Selection} problem was considered in~\cite{KZ15b}, where it was shown that it  is NP-hard for $K=2$. Furthermore, it becomes strongly NP-hard and hard to approximate within $O(\log n)$
(resp. $O(\log K)$)
 when $K$ is unbounded. A maximization version of \textsc{TSt-E Selection} was recently considered in~\cite{corsten2017assortment}.
In this section we will first investigate \textsc{TSt-E Selection} and then use the results obtained to characterize the complexity of more general $\textsc{TSt-CVaR}_{\alpha}~\textsc{Selection}$ problem.

\begin{thm}\label{thm-sel1}
	If the number of scenarios $K$ is unbounded, then \textsc{TSt-E Selection}  is strongly NP-hard and hard to approximate within $O(\log n)$ (resp. $O(\log K)$).
\end{thm}
\begin{proof}
	It follows directly from the result proven in~\cite[Theorem~6]{KZ15b}. It is enough to assume the uniform probability distribution in the resulting scenario set.
\end{proof}
Applying Theorem~\ref{thmehard}, we get the following result:
\begin{cor}\label{cor-sel1}
	If the number of scenarios $K$ is unbounded, then $\textsc{TSt-CVaR}_{\alpha}~\textsc{Selection}$ for any fixed $\alpha\in [0,1)$  is strongly NP-hard and hard to approximate within $O(\log n)$ (resp. $O(\log K)$).
\end{cor}

The next result shows that the problem is polynomially solvable if $K$ is constant.

\begin{thm}
\label{thmselK}
	The \textsc{TSt-E Selection} problem  with $K$ scenarios can be solved in $O(np^{K+1}2^K)$ time, which is polynomial if $K$ is constant.
\end{thm}
\begin{proof}
We use a dynamic programming approach.
	Let $v_i(L,l_1,\dots,l_K)$ be the total cost of an optimal solution for the subset of items $[i]$ under the assumption that $L$ items among $[i]$ are chosen in the first stage and $l_j$ items among $[i]$ are chosen in the second stage under the $j$th scenario, where $i\in\{0,\dots,n\}$ and $L,l_1,\dots,l_K\in \{0,\dots,p\}$.
	Initially $v_0(0,\dots,0)=0$. The values $v_i(L,l_1,\dots,l_K)$ can be recursively computed from the  values $v_{i-1}(L, l_1,\dots,l_K)$ by considering all $2^K+1$ possible placement of the item $i$ (the item $i$ can be chosen in the first stage or distributed in $2^K$ ways over $K$ scenarios). An illustration for $K=2$ is shown in Figure~\ref{fig1}. The optimal solution is contained in the smallest $v_n(L,l_1,\dots,l_K)$ such that $L+l_j=p$ for each $j\in [K]$. The number of different vectors $(L,l_1,\dots,l_K)$ is bounded by $p^{K+1}$. The values of $v_i(L,l_1,\dots,l_K)$ can be computed from $v_{i-1}(L,l_1,\dots,l_K)$ in at most $p^{K+1}2^K$ steps. Hence the overall running time of the algorithm is $O(np^{K+1}2^K)$.
	
	\begin{figure}[ht]
	\centering
	\includegraphics[height=4cm]{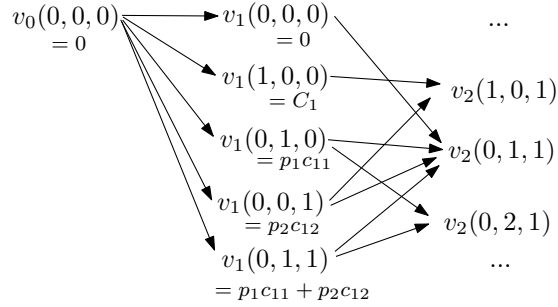}
	\caption{Sample computations for $K=2$, $v_2(1,0,1)=\min\{v_1(1,0,0)+p_2c_{22}, v_1(0,0,1)+C_2\}$, $v_2(0,1,1)=\min\{v_1(0,0,0)+p_1c_{21}+p_2c_{22},v_1(0,1,0)+p_2c_{22}, v_1(0,0,1)+p_1c_{21}, v_1(0,1,1)\}$, {$v_2(0,2,1)= \min\{v_1(0,1,1)+p_1c_{21}, v_1(0,1,0)+p_1c_{21}+p_2c_{22}\}$}.} \label{fig1} 
	\end{figure}
\end{proof}

Theorems~\ref{thmselK} together with Theorem~\ref{thmgen0} imply the following result:
\begin{cor}\label{cor-sel2}
	$\textsc{TSt-CVaR}_{\alpha}~\textsc{Selection}$ is approximable within $\min\{ \frac{1}{{\rm Pr}_{\min}},\frac{1}{1-\alpha}\}$ for constant~$K$ and any fixed $\alpha\in [0,1)$.
\end{cor}

We now provide an LP-based randomized $O(\log n+\log K)$-approximation algorithm for
 $\textsc{TSt-E}~\textsc{Selection}$ and  thus for $\textsc{TSt-CVaR}_{\alpha}~\textsc{Selection}$ for a fixed $\alpha \in [0,1)$
  (see Theorem~\ref{thmgen0}), when $K$ is unbounded. Consider the following linear program:
  \begin{align}
\mathcal{LP}(L):\;&\pmb{C}\pmb{x}+\sum_{j\in [K]}p_j\pmb{c}_j\pmb{y}_j\leq L &\label{scs}\\
                          &  \sum_{i\in [n]} (x_i+y_{ij})=p&   j\in[K],\label{hcs}\\
                          &   x_i+y_{ij}\leq 1&   i\in [n]  j\in[K] \label{hcs0}\\
                          &x_i \in [0,1] &  i\in E(L) \label{hcs1} \\
			&x_i=0 &  i\notin E(L)  \label{hcs2}\\
			&y_{ij} \in [0,1] &i\in E^j(L), j\in [K] \label{hcs3} \\
			&y_{ij}=0 &  i\notin E^j(L), j\in [K]   \label{hcs4}
\end{align}
where $E(L)=\{i\in[n]\,:\, C_i\leq L\}$ and $E^j(L)=\{i\in [n]\,:\, p_jc_{ij}\leq L\}$.
Minimizing $L$ subject to~(\ref{scs})-(\ref{hcs4}), we obtain an LP relaxation of $\textsc{TSt-E}~\textsc{Selection}$.
Let  $L^*$ denote  the smallest value of the parameter~$L$
for which $\mathcal{LP}(L)$ is feasible.
The value $L^*$ is a lower bound on the cost of an optimal solution to $\textsc{TSt-E}~\textsc{Selection}$, and
can be determined in polynomial time by using binary search.  
There is no loss of generality in assuming that 
 $C_i, p_jc_{ij}\in[0,1]$, $i\in[n]$, $j\in [K]$,
 and  $L^*=1$.
We can easily meet this assumption by dividing  the costs of all the items from $E(L)$ and $E^j(L)$, $j\in [K]$,
by~$L^*$. 

We now apply a randomized rounding procedure to convert  a feasible fractional  solution
 $(\pmb{x}^*, \pmb{y}^*)$ 
to $\mathcal{LP}(L^*)$
into a feasible solution to   $\textsc{TSt-E}~\textsc{Selection}$
 (see Algorithm~\ref{alg2sssel}).
 We use $x$-\emph{coin}, i.e. a coin which comes up the head with probability $x\in (0,1]$. 
Let $X$ be the set of the  first stage items chosen and $Y^j$  be the set of the  second stage items chosen
under the $j$th scenario $j\in[K]$. Initially, these sets are empty.
Then,
 for each item~$i\in E(L^*)$, we simply
flip an $x^*_i$-coin $\hat{k}$ times, where $\hat{k}$ will be specified later, and if it comes up heads at least once, then item~$i$  is included in~$X$.  
Similarly,  for every $j\in [K]$ and   
for each item~$i\in E^j(L^*)$, we 
flip an $y^*_{ij}$-coin $\hat{k}$ times, and if it comes up heads at least once, then
 item~$i$  is added  to~$Y^j$. 
If Algorithm~\ref{alg2sssel} outputs a solution such that $|X\cup Y^j|\geq p$ for each $j\in [K]$, then
 the sets $X$ and $Y^j$ can be easily converted  into a feasible solution  to   $\textsc{TSt-E}~\textsc{Selection}$.
 The algorithm fails if $|X\cup Y^j|<p$ for at least one scenario $j$. We will show, however, that this bad event occurs with a small probability.  Note that in this case one can add at most $O(\log n+\log K)$
 items to the first stage set~$X$ from $E(L^*)\setminus X$, if  $E(L^*)\setminus X\not= \emptyset$,
 to repair the feasibility of the solution.
 \begin{algorithm}
\begin{small}
  Find $L^*$ and a feasible solution $(\pmb{x}^*, \pmb{y}^*)$ to $\mathcal{LP}(L^*)$, (\ref{scs})-(\ref{hcs4})\;
  $X\leftarrow\emptyset$ and $Y^j\leftarrow\emptyset$, $j\in [K]$\;
 \tcc{Randomized rounding}
 $\hat{k}\leftarrow \lceil 335\ln n +40\ln 2K\rceil$\; 
 \lForEach(\tcp*[f]{the first stage}){$
 i\in E(L^*)$}{
 flip an $x^*_i$-coin $\hat{k}$ times, if it comes up heads at least once, then
             add~$i$~to  $X$} \label{selsround1}
 \lForEach(\tcp*[f]{the second stage}){$i\in E^j(L^*)$ \KwAND $j\in[K]$}{
 flip an $y^*_{ij}$-coin $\hat{k}$ times, if it comes up heads at least once, then
             add~$i$ to $Y^j$} \label{selsround2}
  \leIf{ $|X\cup Y^j|\geq p$ for each $j\in [K]$} {\Return{ $\{X\cup Y^j\}_{j\in [K]}$}}{\textbf{fail}}  
  \caption{A randomized algorithm for \textsc{TSt-E}~\textsc{Selection}}
 \label{alg2sssel}
\end{small} 
\end{algorithm}

The following analysis of Algorithm~\ref{alg2sssel} is  similar  to that used in~\cite{KZ15b}. 
The total cost of solution~$\{X\cup Y^j\}_{j\in [K]}$ returned  by Algorithm~\ref{alg2sssel}  is established by the following lemma.
 \begin{lem}
 Let $\psi$ be the event that
  $\sum_{i\in X} C_i+\sum_{j\in [K]}\sum_{i\in Y^j}p_jc_{ij}>\hat{k}L^*+(\mathrm{e}-1)\sqrt{\hat{k}L^{*}\ln(2n^2)}$,
  where $\hat{k}=\lceil  335\ln n +40\ln 2K \rceil$. 
  Then $\mathrm{Pr}[\psi]<\frac{1}{2n^2}$.
   \label{lsbou1}
\end{lem}
\begin{proof}
Define  a random variable~$\mathrm{X}_i$ such that $\mathrm{X}_i=1$ if item~$i\in E(L^*)$ is added
 to~$X$;
 and $\mathrm{X}_i=0$ otherwise. Clearly,
$\mathrm{Pr}[\mathrm{X}_i=1]=1-(1-x^*_i)^{\hat{k}}$. 
Fix $j\in [K]$ and define
  a random variable~$\mathrm{Y}_{ij}$ such that $\mathrm{Y}_{ij}=1$ if item~$i\in E^{j}(L^*)$ is added
 to~$Y^{j}$;
 $\mathrm{Y}_{ij}=0$ otherwise and
$\mathrm{Pr}[\mathrm{Y}_{ij}=1]=1-(1-y^*_{ij})^{\hat{k}}$. 
Thus
 \begin{align}
 &\mathbf{E}\left[\sum_{i\in E(L^*)} C_i \mathrm{X}_i+ 
 \sum_{j\in [K]}\sum_{i\in E^{j}(L^*)} p_jc_{ij} \mathrm{Y}_{ij}\right]
   =\sum_{i\in E(L^*)} C_i (1-(1-x^*_i)^{\hat{k}})+ \nonumber\\
  &+ \sum_{j\in [K]}\sum_{i\in E^{j}(L^*)} p_jc_{ij} (1-(1-y^*_{ij})^{\hat{k}})
             \leq \hat{k}\left(\sum_{i\in E(L^*)} C_i x^*_i +
              \sum_{j\in [K]}\sum_{i\in E^{j}(L^*)} p_j c_{ij}y^*_{ij} \right)\leq \hat{k}L^*, \label{exspcoub}
 \end{align}
 where the  inequality in~(\ref{exspcoub}) is due to the fact that
 $1-(1-z)^{\hat{k}}\leq  \hat{k}z$ for $\hat{k}\geq 1$ and $z\in (0,1]$.
As $C_i,p_jc_{ij}\in [0,1]$, we can then use the 
 Chernoff-Hoeffding bound 
(see Theorem~1 and inequality (1.13) from \cite{R88} for $D(\hat{k}L^*,1/(2n^2))$, $\hat{k}L^*> \ln(2n^2)$) to
show that
$\mathrm{Pr}\left[\psi\right]< \frac{1}{2n^2}$,
 which proves the lemma.
 \end{proof}
We now consider the feasibility of the solution output $\{X\cup Y^j\}_{j\in [K]}$. 
 Notice first that Steps~\ref{selsround1} and~\ref{selsround2}  can be seen as performing $\hat{k}$ rounds independently.
 Namely, in each round~$k$, $k\in[\hat{k}]$,
we flip an $x^*_i$-coin for each item~$i\in E(L^*)$ and include~$i$ into~$X$ when it comes up the head,  
and for every $j\in [K]$ and   
for each item~$i\in E^j(L^*)$, we 
flip an $y^*_{ij}$-coin and if it comes up the head, then
item~$i$  is added  to~$Y^j$.                       
Let $X_k$ and $Y^j_k$ be the sets of items selected in the first and second stage under $j\in[K]$, respectively, after $k$ rounds.  Define $E^j_k=[n]\setminus (X_k\cup Y^j_k)$ and $|E^j_k|=N^j_k$. Initially, $E^{j}_0=[n]$, $N^j_0=n$. 
Let $P^j_k$ denote the number of items remaining for selection out of the set $E^{j}_k$  under scenario~$j$ 
after the $k$th  round. Initially $P^j_0=p$.
We say that a round~$k$ is ``successful''  if
either $P^j_{k-1}=0$  or
$P^j_{k}<0.98P^j_{k-1}$;
otherwise, it is ``failure''.
The following two lemmas are a slight modification of the ones given 
in~\cite[Lemmas 7 and 8]{KZ15b}. We include their proofs for completeness in Appendix~\ref{dod}.
\begin{lem}
Fix scenario $j\in [K]$. 
Then for every~$k$
the conditional probability that round~$k$ is  
``successful'', given any set of items~$E^{j}_{k-1}$ 
and number~$P^j_{k-1}$,
is at least $1/5$.
\label{lsalon}
\end{lem}
\begin{lem}
Let $\xi^j$ be the event that $P^j_{\hat{k}}\geq 1$. Then
$\mathrm{Pr}[\xi^j]<\frac{1}{2Kn^2}$, provided that
 $\hat{k}= \lceil 335\ln n + 40\ln 2K\rceil$.
  \label{lsbou2}
 \end{lem}
Accordingly, the union bound  gives 
$\Pr[\psi\cup \xi^1\cup \dots\cup \xi^K]< 1/n^2$
(see Lemmas~\ref{lsbou1}  and~\ref{lsbou2}).
Hence,
after  $\hat{k}= \lceil 335\ln n +40\ln 2K \rceil$ rounds, Algorithm~\ref{alg2sssel}
yields a feasible solution $\{X\cup Y^j\}_{j\in [K]}$ with the total cost of
$O(\ln n +\ln K)L^*$
with probability at least $1-\frac{1}{n^2}$. We
thus get the following theorem.
\begin{thm}
There is a randomized approximation algorithm for   \textsc{TSt-E}~\textsc{Selection}
that yields  an $O(\log n +\log K)$-approximate solution with high probability.
\label{tserand}
\end{thm}
Algorithm~\ref{alg2sssel}  is tight  up to a constant factor, when $K=\mathrm{poly}(n)$
(see  Theorem~\ref{thm-sel1}).
From the above and Theorem~\ref{thmgen0} we immediately get
 the following result:
\begin{cor}\label{tsecvrand}
There is a randomized approximation algorithm for   $\textsc{TSt-CVaR}_{\alpha}~\textsc{Selection}$
that yields  an $O(\log n +\log K)$-approximate solution with high probability for any fixed $\alpha\in [0,1)$.
\end{cor}


\section{Complexity of two-stage network problems}
\label{secnetworks}

In this section we consider a class of basic network problems, in which $\mathcal{X}$ is a set of characteristic vectors of some objects in a given graph such as paths, spanning trees and matchings.


\subsection{Shortest path problem}
\label{secsp}

Let $G=(V,A)$ be a directed graph. In the \textsc{Shortest Path} problem, $\mathcal{X}$ is the set of characteristic vectors of the simple $s-t$ paths in $G$. The sample problem presented in Figure~\ref{fig3}, shows that we can improve the quality of the solution by extending $\mathcal{X}$ to $\overline{\mathcal{X}}$. So, it is interesting to explore to computational properties of the problem with both $\mathcal{X}$ and $\overline{\mathcal{X}}$.

\begin{thm}
\label{thmsp1}
	\textsc{TSt-R~Shortest Path} is NP-hard for $K=2$ and strongly NP-hard for unbounded $K$, even if $|\mathcal{X}|=|\overline{\mathcal{X}}|=1$, i.e. if there is only one feasible solution.
\end{thm}
\begin{proof}
	Consider the network (chain) shown in Figure~\ref{fig2}. In this instance $\mathcal{X}=\overline{\mathcal{X}}$ and both sets contain only one solution. The problem is now to decide,  for each arc, whether to choose it in the first or in the second stage.  This problem is equivalent to \textsc{TSt-R RS} and Theorems~\ref{thmrs1} and~\ref{thmrs2} immediately imply the result.
	
	\begin{figure}[ht]
	\centering
	\includegraphics{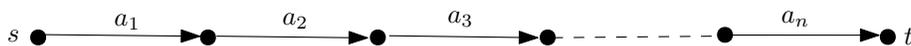}
	\caption{Illustration of the proof of Theorem~\ref{thmsp1}.} \label{fig2} 
	\end{figure}
\end{proof}

\begin{thm}
\label{thmsp3}
	\textsc{TSt-E} and \textsc{TSt-R Shortest Path} with set $\mathcal{X}$ are strongly NP-hard and not at all approximable even if $K=2$.
\end{thm}
\begin{proof}
	Consider the following \textsc{Hamiltonian Path} problem, which is known to be NP-complete~\cite{GJ79}. We are given a directed graph $G=(V,A)$ and ask if there is a directed path from $v_1$ to $v_n$ in $G$ that visits each node in $V$ exactly once. An example instance with four nodes is given in Figure~\ref{fig8}a. Given an instance of \textsc{Hamiltonian Path} we build the corresponding instance of \textsc{TSt-E (R) Shortest Path} as follows. The network $G'=(V',A')$ contains two nodes $v_i$ and $v_i'$ for each $v_i\in V$. The set of arcs $A'$ contains the \emph{forward arcs} $(v_i, v_i')$ for each $v_i\in V$ and \emph{backward arcs} $(v_i',v_j)$ for each $i\neq j$ (see Figure~\ref{fig8}b). The first stage costs of the forward arcs are~0 and the first stage costs of the backward arcs are are~1. We now create two second-stage cost scenarios as follows. In scenario $\pmb{c}_1$ the costs of the arcs $(v_i', v_{i+1})$, $i\in [n-1]$, are 0 and the cost of all the remaining arc are~1 (see Figure~\ref{fig8}c). In   scenario $\pmb{c}_2$, we fix the cost of $(v_i', v_j)$ to~0 for each $(v_i, v_j)\in A$, and the costs of all the remaining arcs are~1 (see Figure~\ref{fig8}d).  We set $s=v_1$, $t=v_n$ and fix any probability distribution, $p_1,p_2>0$, in the scenario set.

	\begin{figure}[ht]
	\centering
	\includegraphics[height=5cm]{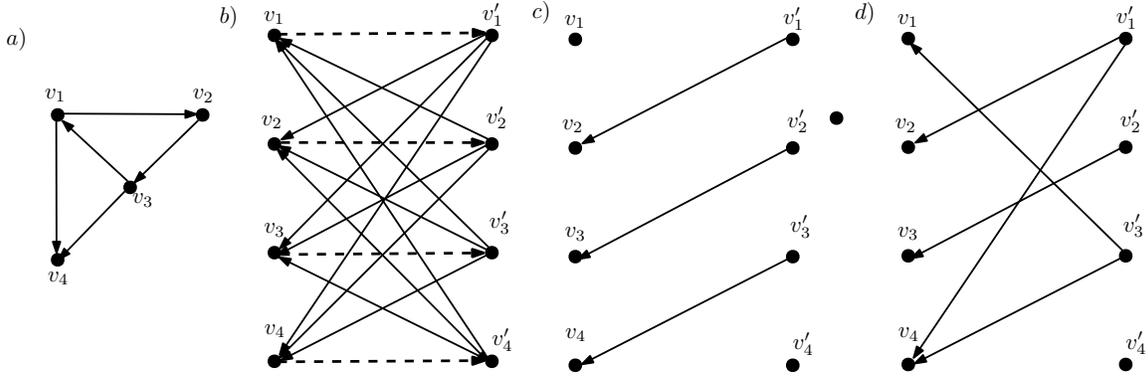}
	\caption{Illustration of the proof of Theorem~\ref{thmsp3}.} \label{fig8} 
	\end{figure}
	
We will show that the answer to \textsc{Hamiltonian Path} is yes if and only if there is a solution to \textsc{TSt-E (R) Shortest Path} with set $\mathcal{X}$ with the expected (maximum) cost equal to~0.
Assume that there is a Hamiltonian path $(v_{\sigma(1)}, v_{\sigma(2)}, \dots, v_{\sigma(n)})$ in $G$, where $v_{\sigma(1)}=v_1$ and $v_{\sigma(n)}=v_n$. We choose all forward arcs in the first stage. Under scenario $\pmb{c}_1$ we build the $s-t$ path by choosing all the arcs with 0 second-stage costs. Under $\pmb{c}_2$ we form the $s-t$ path by adding the arcs $(v_{\sigma(i)}', v_{\sigma(i+1)})$, $i\in [n-1]$. By the construction, the second stage costs of all these arcs are~0. Hence the maximum (expected) cost of the solution is~0. Assume that there is a solution to \textsc{TSt-E (R) Shortest Path} with the expected (maximum) cost equal to~0. Using $\pmb{c}_1$, one can observe that all the forward arcs must be chosen in the first stage (there is a unique $s-t$ path with 0 cost under the first scenario and this path contains all forward arcs). Now, using $\pmb{c}_2$, one has to form a simple $s-t$ path with~0 costs by adding backward arcs with~0 costs to all forward arcs (recall that no redundant arcs are allowed in set $\mathcal{X}$). This path has a form of $(v_{\sigma(1)}, v_{\sigma(1)}',v_{\sigma(2)},v_{\sigma(2)}',v_{\sigma(3)},\dots,v_{\sigma(n)},v_{\sigma(n)}')$, where $v_{\sigma(1)}=v_1$ and $v_{\sigma(n)}=v_n$. We can now observe that $(v_{\sigma(1)}, v_{\sigma(2)}, \dots, v_{\sigma(n)})$ is a Hamiltonian path in $G$.

\end{proof}

\begin{cor}
	$\textsc{TSt-CVaR}_\alpha~\textsc{Shortest Path}$ with set $\mathcal{X}$ is  is strongly NP-hard and not at all approximable for each fixed $\alpha\in [0,1)$, even if $K=2$.
\end{cor}
\begin{proof}
	The proof follows directly from the reduction in the proof of Theorem~\ref{thmsp3}.
\end{proof}

\begin{thm}
\label{thmsp4}
	\textsc{TSt-E} and \textsc{TSt-R Shortest Path} with set $\overline{\mathcal{X}}$ are strongly NP-hard, even if $K=2$.	
\end{thm}
\begin{proof}
	
	Consider the following \textsc{Sat} problem, which is known to be NP-complete~\cite{GJ79}. We are given $n$ boolean variables $x_1\dots,x_n$ and a set  of clauses $\mathcal{C}_1,\dots,\mathcal{C}_m$, where each clause is a disjunction of some literals  (variables or their negations). We ask if there is a truth assignment to the variables which satisfies all the clauses. Given an instance of \textsc{Sat} we create the corresponding instance of \textsc{TSt-E (R) Shortest Path} as follows. For each variable $x_i$, $i\in [n]$, we create a component shown in Figure~\ref{fig9a}. The arc $a_{ij}$ corresponds to the case in which $x_i=1$ satisfies clause $\mathcal{C}_j$ and the arc $b_{ij}$ corresponds to the case in which $x_i=0$ satisfies $\mathcal{C}_j$. Notice that there is also a number of \emph{dashed arcs} in the component.
	 We merge the components by identifying $s_{i+1}$ with $t_{i}$ for $i\in [n-1]$ and we set $s=s_1$, $t=t_n$. Consider now clause $\mathcal{C}_j$. We add nodes $v_{ja}$ and $v_{jb}$ to the network. If literal $x_i$ appears in $\mathcal{C}_j$ we add two arcs so that there is a path from $v_{ja}$ to $v_{jb}$ using the arc $a_{ij}$. Similarly, if literal $\overline{x}_i$ appears in $\mathcal{C}_j$, then we add two arcs so that there is a path from $v_{ja}$ to $v_{jb}$ using the arc $b_{ij}$. We add arcs $(v_{jb},v_{j+1,a})$ for $j\in [n-1]$ (see the example shown in Figure~\ref{fig9}). We complete the construction of the network by adding the arcs $(s, v_{1a})$ and $(v_{mb},t)$. All the arcs outside the components are called \emph{clause arcs}.
	
	\begin{figure}[ht]
	\centering
	\includegraphics[height=2cm]{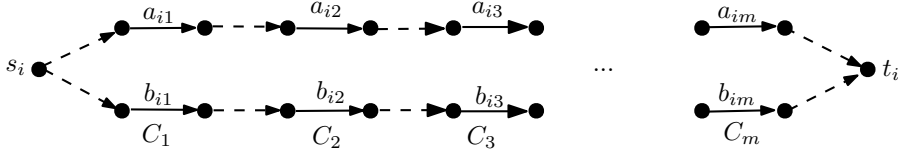}
	\caption{A component corresponding to variable $x_i$.} \label{fig9a} 
	\end{figure}

	\begin{figure}[ht]
	\centering
	\includegraphics[height=5cm]{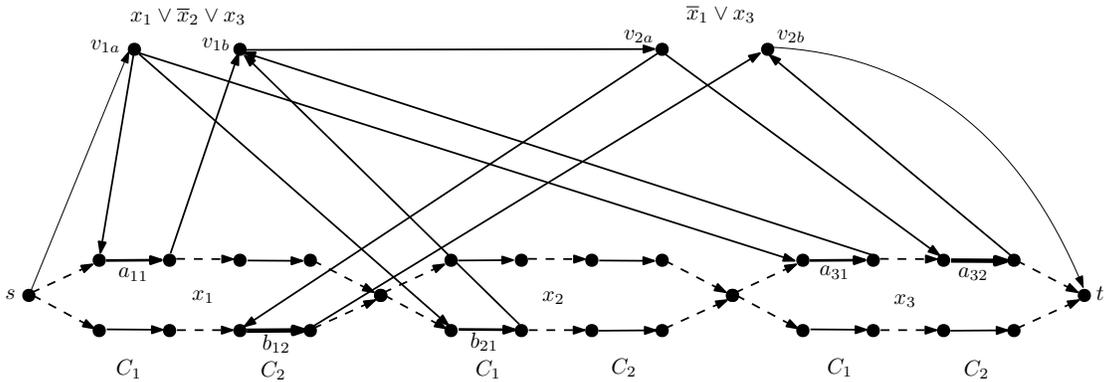}
	\caption{Illustration of the proof of Theorem~\ref{thmsp4}.} \label{fig9} 
	\end{figure}

The first stage costs of the arcs $a_{ij}$ and $b_{ij}$ are set to~1 and the first stage costs of the remaining arcs are set to $M=2nm+2$. We now form two second-stage cost scenarios as follows. In the first scenario $\pmb{c}_1$ the costs of all dashed arcs in the components are set to~0 and the costs of all the remaining arcs are set to~$M$. In the second scenario  $\pmb{c}_2$, the costs of all clause arcs are~0 and the costs of all the remaining arcs are equal to~$M$. We fix the probabilities of both scenarios to $0.5$. We will show that the answer to \textsc{Sat} is yes if and only if there is a solution to \textsc{TSt-E (R) Shortest Path} with the expected (maximum) cost at most $mn$.

Assume that the answer to \textsc{Sat} is yes, so there is a truth assignment to the variables which satisfies all the clauses. If $x_i=1$, then we choose in the first stage all arcs $a_{ij}$, $j\in [m]$ and if $x_i=0$, then we choose in the first stage all arcs $b_{ij}$, $j\in [m]$. The overall first stage solution cost is thus $mn$. By choosing all dashed arcs under $\pmb{c}_1$ we connect $s$ and $t$, and the cost of the recourse action under $\pmb{c}_1$ is~0. Because the truth assignment satisfies all clauses, we can also connect $s$ and $t$ under $\pmb{c}_2$ by adding a number of clause arcs. The cost of this recourse action is also~0. Hence the expected (maximum) cost of the constructed solution is $mn$.
Assume that there is a solution to \textsc{TSt-E (R) Shortest Path} with the expected (maximum) cost at most $mn$. Observe that in the first stage, for each $i\in [n]$, we have to choose either all arcs $a_{ij}$ or all arcs $b_{ij}$, $j\in [m]$. Otherwise, the cost of the recourse action under $\pmb{c}_1$ will be at least $2mn+2$ and the expected (maximum) cot of the solution is greater than $mn$.  Because the first stage cost of the solution is now exactly $mn$, the cost of the recourse action under $\pmb{c}_2$ must be~0. So, we must be able to add a number of clause arcs so that $s$ and $t$ are connected. We can observe that his implies that there is a truth assignment to the variables which satisfies all clauses.

\end{proof}

\begin{cor}
	$\textsc{TSt-CVaR}_\alpha~\textsc{Shortest Path}$ with set $\overline{\mathcal{X}}$ is  is strongly NP-hard for each fixed $\alpha\in [0,1)$, even if $K=2$.
\end{cor}
\begin{proof}
	The proof follows directly from the reduction in the proof of Theorem~\ref{thmsp4}.
\end{proof}

\begin{thm}
\label{thmsp5}
	\textsc{TSt-E} and \textsc{TSt-R Shortest path} with set $\overline{\mathcal{X}}$ and unbounded $K$ are strongly NP-hard and hard to approximate within $O(\log K)$ even if the input graph is series-parallel. 
\end{thm}
\begin{proof}
	We will show a cost preserving reduction from the \textsc{Min-Set Cover} problem (see the proof of Theorem~\ref{thmrs2}). The inapproximability of \textsc{TSt-E (R) Shortest Path} follows then form the inapproximability of \textsc{Min-Set Cover} (see \cite{feige1998threshold}).
	We will assume w.l.o.g that each element $u_i\in \mathcal{U}$ belongs to at least one set $A_j\in \mathcal{A}$, so there is a trivial cover of size $|\mathcal{A}|=m$.  Given an instance $(\mathcal{U},\mathcal{A})$ of \textsc{Min-Set Cover} we build a network $G=(V,A)$, shown in Figure~\ref{fig7}. For each set $A_j\in \mathcal{A}$, $j\in [m]$, this graph contains an $s-t$ path consisting of two arcs $a_j$ and $a_j'$, respectively (see Figure~\ref{fig7}). The first stage costs of $a_j$ are 1 and the first stage costs of $a_j'$ are $m+1$ for all $j\in [m]$. For each element $u_i$, $i\in [n]$, we create a second stage cost scenario $\pmb{c}_i$ as follows. The cost of $a_j$, $j\in [m]$, is equal to $m+1$. The cost of $a_j'$ is equal to~0 if $u_i\in A_j$ and $m+1$, otherwise. Notice that $K=n$. We can fix any probability distribution in the resulting scenario set. We will show that there is a cover $D$ of size $L$ if and only if there is a solution to \textsc{TSt-E (R) Shortest path} with the expected (maximum) cost equal to $L$.
	
	\begin{figure}[ht]
	\centering
	\includegraphics[height=4cm]{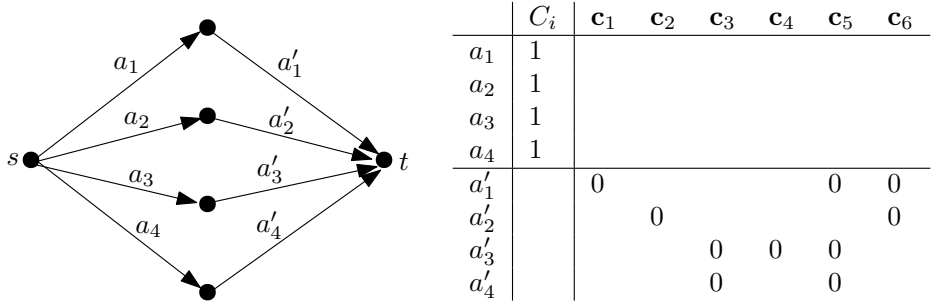}
	\caption{An instance of the problem for $\mathcal{U}=\{u_1,\dots,u_6\}$ and $\mathcal{A}=\{\{u_1,u_5,u_6\}, \{u_2,u_6\}, \{u_3,u_4,u_5\}, \{u_3,u_5\}\}$. The empty entries in the table contain the value $m+1$.} \label{fig7} 
	\end{figure}	

Assume that there is a cover $D$ such that $|D|=L$. In the first stage we choose the arcs $a_j$ for each $A_j\in D$.
Consider scenario $\pmb{c}_i$ corresponding to element $u_i$, $i\in [n]$. Because $D$ is a cover, there is at least one arc $a_j'$ with 0 cost under $\pmb{c}_i$, which added to the arcs selected in the first stage connects $s$ and $t$.  Because the cost of the recourse action is~0 under each scenario the expected (maximum) cost of the constructed solution is $L$. 

Assume that the expected (maximum) cost of the solution to  \textsc{TSt-E (R) Shortest path} is equal to $L\leq m$. Observe, that the cost of the recourse action under each scenario must be~0 and so exactly $L$ arcs among $a_j$, $j\in [m]$, are chosen in the first stage. It is easy to observe that the sets $A_j$ corresponding to the chosen arcs form a cover $D$ of size $L$.

\end{proof}

\begin{cor}
	$\textsc{TSt-CVaR}_\alpha~\textsc{Shortest Path}$ with set $\overline{\mathcal{X}}$, unbounded $K$  and any fixed $\alpha\in [0,1)$ is strongly NP-hard and hard to approximate within $O(\log K)$ even if the input graph is series-parallel.
\end{cor}
\begin{proof}
	The proof follows directly from Theorem~\ref{thmsp5} and Theorem~\ref{thmehard}.
\end{proof}	

\begin{thm}\label{thmsp6}
	\textsc{TSt-E Shortest path} with set $\mathcal{X}$ is polynomially solvable in series-parallel graphs.
\end{thm}
\begin{proof}
We sketch an inductive proof. Given a series-parallel graph, we first find in polynomial time its decomposition tree into parallel and sequential components~\cite{VTL82}. Note that the following two cases can be solved in polynomial time: a graph with only two nodes $s$ and $t$, which are connected by two parallel arcs, and a graph consisting of three nodes $\{s,v,t\}$ and two sequential arcs.
Now let two series-parallel graphs $G_1$ and $G_2$ be given, and consider their parallel combination. This can be considered as a problem of type \textsc{TSt-E RS}, which can be solved in polynomial time (see Observation~\ref{obs1}) by choosing the cheaper of the two sub-solutions. For two series-parallel graph $G_1$ and $G_2$ which are combined sequentially, we combine the respective solutions and their objective values add up.
\end{proof}

Note that using the set $\mathcal{X}$ is required in Theorem~\ref{thmsp6} for choosing the cheaper of two solutions in the case of parallel combinations. Otherwise, for set $\overline{\mathcal{X}}$, a partial combination of solutions would be possible (and our method cannot be applied). In fact, the reduction in the proof of Theorem~\ref{thmsp5} shows that the problem with set $\overline{\mathcal{X}}$ is hard for series-parallel graphs.

\begin{cor}
        $\textsc{TSt-CVaR}_\alpha~\textsc{Shortest Path}$ for series-parallel graphs with set $\mathcal{X}$ is approximable within  $\min\left \{\frac{1}{{\Pr}_{\min}},\frac{1}{1-\alpha} \right\}$ .
\end{cor}
\begin{proof}
	A direct consequence of Theorem~\ref{thmsp6} and Theorem~\ref{thmgen0}.
\end{proof}

We now consider another variant of the two-stage shortest path problem. Given a network $G=(V,A)$, suppose that in the first stage we must choose a set of arcs which form a path from $s$ to some node $v_i\in V$ (if $v_i=s$, then we stay at $s$ in the first stage). Note that this is an additional restriction imposed on $\mathcal{X}'$.
 This problem corresponds, for example, to the situation in which we travel to some point and continue the trip after a second-stage scenario reveals. In order to distinguish the problem from the previously discussed, we will call it \textsc{Connectivity}. 

\begin{obs}
\label{obs2}
	$\textsc{TSt-CVaR}_\alpha~\textsc{Connectivity}$ in acyclic graphs is polynomially solvable for any $\alpha\in [0,1)$.
\end{obs}
\begin{proof}
	Observe that the problem reduces to finding a node $v_i^*\in V$, which is reached in the first stage. This can be done by computing for each $v_i\in V$, the cost of a shortest path $\pmb{C}^*_i$ from $s$ to $v_i$ with respect to the first stage costs and the cost $\pmb{c}^*_{ij}$ of a shortest path from $v_i$ to $t$ under scenario $\pmb{c}_j$, $j\in [K]$. Let ${\rm Y}_i$ be a random variable taking the value $\pmb{c}^*_{ij}$ with probability $p_j$. We can now compute the value of $\pmb{C}^*_i+{\bf CVaR}_{\alpha}[{\rm Y}_i]$ for each $v_i$ and choose the node $v^*_i\in V$, for which this value is minimal. All the computations can be done in polynomial time. Observe that the resulting path is simple, as the input graph is acyclic.
\end{proof}

The situation is more complex if the network is not acyclic. In this case, the resulting path need not to be simple because some arcs chosen in the first stage can also be chosen in the second stage. If the resulting path must be simple and the input graph is not acyclic, then the algorithm described in Observation~\ref{obs2} does not work and the complexity of the problem remains open.

\subsection{Minimum spanning tree}

Let $G=(V,E)$, $|V|=n$, $|E|=m$, be an undirected network. In the \textsc{Spanning Tree} problem, $\mathcal{X}$ contains characteristic vectors of the spanning trees in $G$. We first note that the problem with set $\overline{\mathcal{X}}$ is equivalent to the problem with $\mathcal{X}$, i.e. it is not profitable to choose more edges  than necessary in the first stage. Observe, that in this case the set of edges chosen in the first stage contains a cycle $C$. Let $f\in C$ be any edge on this cycle. Let $T_i$ be the set of edges chosen in the first and second stage under scenario $i\in [K]$. Of course, $C\subseteq T_i$ and $T_i$ contains a spanning tree, so the subgraph induced by $T_i$ is connected. If we remove $f$ from $T_i$ the resulting subgraph is still connected and $T_i\setminus \{f\}$ also contains a spanning tree for each $i\in [K]$.  In consequence, there is  an optimal solution that does not use $f$ in the first stage. Using this argument repetitively, one can see that it is enough to choose an acyclic subset of edges in the first stage.

The \textsc{TSt-E Spanning Tree} and \textsc{TSt-R Spanning Tree} problems were investigated in~\cite{DRM05}, \cite{FFK06} and~\cite{KZ11}. 
It has been shown 
in~\cite{FFK06} and~\cite{KZ11}   that both problems are strongly NP-hard and hard to approximate within $O(\log K)$ (resp. $O(\log n)$) in complete graphs
 for unbounded $K$. 
 Hence and from Theorem~\ref{thmehard} we get immediately get the following result:
\begin{thm}
The  $\textsc{TSt-CVaR}_{\alpha}~\textsc{Spanning Tree}$  problem is 
 strongly NP-hard and
hard to approximate within  $O(\log K)$ (resp. $O(\log n)$)
 in complete graphs
for unbounded $K$ and any fixed $\alpha \in [0,1)$.
\end{thm}
It turns out that that \textsc{TSt-R Spanning Tree} remains hard for constant $K$ in very simple graphs, consisting of only
series edges.
\begin{thm}
\label{thmst1}
	The \textsc{TSt-R~Spanning Tree} problem is NP-hard for $K=2$ and strongly NP-hard for unbounded $K$, even if $|\mathcal{X}|=1$, i.e. if there is only one feasible solution.
\end{thm}

\begin{proof}
The proof is the same as the proof of Theorem~\ref{thmsp1}. Observe, that the network in Figure~\ref{fig2} is a spanning tree (after ignoring the arc directions).
\end{proof}
\begin{cor}
	The $\textsc{TSt-CVaR}_{\alpha}~\textsc{Spanning Tree}$ problem is NP-hard for each $\alpha \in [0.5,1)$.
	\label{corst1}
\end{cor}
\begin{proof}
	Analogous to the proof of Corollary~\ref{cors1}.
\end{proof}
The complexity of $\textsc{TSt-CVaR}_{\alpha}~\textsc{Spanning Tree}$ for $\alpha\in [0,1)$ and constant $K$ is an interesting open problem.

We now present positive results for \textsc{TSt-R Spanning Tree}, \textsc{TSt-E Spanning Tree}
and  $\textsc{TSt-CVaR}_{\alpha}~\textsc{Spanning Tree}$, when $K$ is unbounded.
In~\cite{KZ11}, a randomized $O(\log^2 n)$-approximation algorithm for \textsc{TSt-R Spanning Tree} was constructed.
In~\cite{DRM05} a randomized algorithm was proposed that outputs a solution
whose expected  cost is within  $O(\log n + \log K)$ of the cost of an optimal
solution to  \textsc{TSt-E Spanning Tree}.
We now improve those results and
provide two randomized  $O(\log n + \log K)$-approximation algorithms for 
\textsc{TSt-R Spanning Tree}  and  \textsc{TSt-E Spanning Tree}.
They are similar in spirit to the randomized algorithm  for   \textsc{TSt-E Selection}
(see Section~\ref{secsel}).
 We present only the algorithm for \textsc{TSt-R Spanning Tree},
  the corresponding algorithm for  \textsc{TSt-E Spanning Tree} is the same and its analysis 
 goes in similar manner to that for the robust counterpart.
Consider the following linear program:
\begin{align}
\mathcal{LP}(L):\;&\pmb{C}\pmb{x}+\pmb{c}_j\pmb{y}_j\leq L &   j\in[K] \label{sc}\\
                          &  \sum_{e\in\delta(S)} x_e+y_{e\,j}\geq 1&  \emptyset\not=S\subset V,  j\in[K] \label{hc}\\
                          &x_e \in [0,1] &  e\in E(L)  \label{hc1} \\
			&x_e=0 &  e\notin E(L)  \label{hc2}\\
			&y_{e\,j} \in [0,1] &  e\in E^j(L), j\in [K] \label{hc3} \\
			&y_{e\,j}=0 &  e\notin E^j(L), j\in [K]  \label{hc4}
\end{align}
where $\delta(S)$  is the cut-set determined by node set $S$, i.e.
$\delta(S)=\{ e=\{k,l\} \in E\, :\, k\in S, l\in V\setminus S\}$, 
$L>0$ is a fixed paremeter and
$E(L)=\{e\in E:  C_e\leq L\} \subseteq E$ and
$E^j(L)=\{e\in E:  c_{e\,j}\leq L\} \subseteq E$, $j\in[K]$.
The constraints~(\ref{hc}), describing set~$\mathcal{X}$,
are  the core  of the \emph{cut-set formulation} for the minimum spanning 
tree~\cite{TLMLAW95}.
Minimizing $L$ subject to~(\ref{sc})-(\ref{hc4}) gives an LP relaxation of  a 
mixed integer programming formulation, with integral constraints (see (\ref{hc1}) and (\ref{hc3})),
for \textsc{TSt-R Spanning Tree}. Hence the smallest value of the parameter~$L$, denoted by~$L^*$, 
provides a lower bound on  the cost of an optimal solution to \textsc{TSt-R Spanning Tree}.
We use  binary search to determine~$L^*$. 
The polynomial time solvability of  $\mathcal{LP}(L)$ follows from an efficient  polynomial time separation, based on 
the min-cut problem (see, e.g.,~\cite{TLMLAW95}).

Similarly as  in Section~\ref{secsel} for the  \textsc{TSt-E Selection} problem, we can assume that
 all the edge costs  are such that~$C_e, c_{e\\,j}\in[0,1]$, $e\in E$, $j\in [K]$,
 and thus $L^*=1$. Here
we can use  again a coin to transform a feasible fractional solution $(\pmb{x}^*, \pmb{y}^*)$ 
to $\mathcal{LP}(L^*)$
into a feasible solution to  \textsc{TSt-R Spanning Tree} (see Algorithm~\ref{alg2sstr}).
Namely,
we start with $K$ forest $T^j=(V,F\cup F^j)$, $j\in [K]$, where 
  $F$ is the set of edges chosen in the first stage and $F^j$ is the set of edges chosen in the second stage under
  the $j$th scenario, $j\in [K]$.
Then,
 for each edge~$e\in E(L^*)$, we 
flip an $x^*_e$-coin $\hat{k}$ times and if it comes up heads at least once, then edge~$e$  is included in~$F$
and    for every $j\in [K]$ and   
for each edge~$e\in E^j(L^*)$, we 
flip an $y^*_{e\,j}$-coin $\hat{k}$ times, and if it comes up heads at least once, then
 edge~$e$  is added  to~$F^j$.                       
If all the resulting graphs~$T^j$, $j\in [K]$,  are connected, then 
we can easily construct a feasible solution to  \textsc{TSt-R Spanning Tree}. Otherwise,
 the algorithm fails. Observe that
 in order to repair  $\{T^j\}_{j\in [K]}$  one can try to include to it at most $O(\log n+\log K)$
 edges    from the set 
 $\left(E(L^*)\cup \bigcup_{j\in [K]} E^j(L^*)\right)
 \setminus \left( F \cup \bigcup_{j\in [K]} F^j\right)$ if this set is not empty.
\begin{algorithm}
\begin{small}
  Find $L^*$ and a feasible solution $(\pmb{x}^*, \pmb{y}^*)$ to $\mathcal{LP}(L^*)$, (\ref{sc})-(\ref{hc4} \label{slp})\;
 Let $T^j=(V,F\cup F^j)$, where $F\leftarrow\emptyset$ and $F^j\leftarrow\emptyset$, $j\in [K]$\;
 \tcc{Randomized rounding}
 $\hat{k}\leftarrow \lceil 40\ln n +16\ln K\rceil$\; 
 \lForEach(\tcp*[f]{the first stage}){$e\in E(L^*)$}{
 flip an $x^*_e$-coin $\hat{k}$ times, if it comes up heads at least once, then
             add~$e$~to  $F$} \label{sround1}
 \lForEach(\tcp*[f]{the second stage}){$e\in E^j(L^*)$ \KwAND $j\in[K]$}{
 flip an $y^*_{e\,j}$-coin $\hat{k}$ times, if it comes up heads at least once, then
             add~$e$ to $F^j$} \label{sround2}            
\leIf{all $T^j$, $j\in [K]$, are connected} {\Return{ $\{T^j\}_{j\in [K]}$}}{\textbf{fail}}             
  \caption{A randomized algorithm for \textsc{TSt-R Spanning Tree}}
 \label{alg2sstr}
\end{small} 
\end{algorithm}

The following lemma that provides
the cost  of edges in $T^j=(V,F\cup F^j)$ under scenario~$j$, chosen by Algorithm~\ref{alg2sstr},
 may be proved in much the same way as Lemma~\ref{lsbou1}.
 \begin{lem}
 Let $\psi^j$ be the event that
 $\sum_{e\in F} C_e+\sum_{e\in F^j}c_{e\,j}>\hat{k}L^*+(\mathrm{e}-1)\sqrt{\hat{k}L^{*}\ln(2(nK)^2)}$,
   under scenario
$j\in [K]$,  where $\hat{k}=\lceil 40\ln n+16\ln K \rceil$. Then
  $\mathrm{Pr}\left[\psi^j\right]< \frac{1}{2(nK)^2}$.
   \label{lbou1}
\end{lem}
We now focus on feasibility of  $\{T^j\}_{j\in [K]}$ and
  estimate the probability of the event that all the graphs~$T^j=(V,F\cup F^j)$, $j\in[K]$, built in
 Steps~\ref{sround1} and~\ref{sround2} of Algorithm~\ref{alg2sstr}, are connected.
 These steps are equivalent to  performing $\hat{k}$ rounds independently.
 Indeed, in each round~$k$, $k\in[\hat{k}]$,
we flip an $x^*_e$-coin for each edge~$e\in E(L^*)$ and include~$e$ into~$F$ when it comes up head,  and for every $j\in [K]$ and   
for each edge~$e\in E^j(L^*)$, we 
flip an $y^*_{e\,j}$-coin and if it comes up head, then
edge~$e$  is added  to~$F^j$.                       
The further analysis is adapted from~\cite{A95} and~\cite{DRM05}. Let $T^j_k=(V,F_k\cup F^j_k)$ be the graph obtained from $T^j_{k-1}$, $j\in[K]$,
 after  the~$k$th round, $k\in[\hat{k}]$.
Initially, $T^j_0$ has no edges.
Let $C^j_k$ stand for the number of connected components  of~$T^j_k$. Obviously,
$C^j_0=n$.
We say that  round~$k$ is ``successful'' if
either $C^j_{k-1}=1$ ($T^j_{k-1}$ is connected) or
$C^j_{k}<0.9C^j_{k-1}$.
\begin{lem}[\cite{A95}]
Fix scenario $j\in [K]$.
Assume that for every connected component~$D^j$ of~$T^j_{k-1}$,
the sum of probabilities associated to edges  that connect   nodes of~$D^j$ to nodes 
outside~$D^j$ is at least~1. Then
for every~$k$, the conditional probability that round~$k$ is  
``successful'', given any set of components in $T^j_{k-1}$,
is at least $1/2$.
\label{lalon}
\end{lem}
Obviously, in our case the assumption of Lemma~\ref{lalon} is satisfied, which is due to
the form of constraints~(\ref{hc}) in the linear program~$\mathcal{LP}(L)$.
The above lemma is exploited in the proof 
following result, which specifies the number of rounds~$\hat{k}$.
\begin{lem}[\cite{DRM05}]
Let $\xi^j$ be the event that $C^j_{\hat{k}}\geq 2$
($T^j_{\hat{k}}$ is not connected). Then
$\mathrm{Pr}[\xi^j]<\frac{1}{2(nK)^2}$, provided that
 $\hat{k}= \lceil 40\ln n + 16\ln K\rceil$.
  \label{lbou2}
 \end{lem}
\begin{proof}
The proof goes in similar manner
  to the one of Lemma~\ref{lsbou2} (see Appendix~\ref{dod}).
\end{proof}
Lemmas~\ref{lbou1}  and~\ref{lbou2}  and the union bound  yield
$\Pr[\psi^1\cup \dots\cup\psi^K\cup \xi^1\cup \dots\cup \xi^K]< 1/(Kn^2)$.
Thus with probability at least $1-\frac{1}{Kn^2}$ 
 Algorithm~\ref{alg2sstr} outputs (after  $\hat{k}= \lceil 40\ln n +16\ln K \rceil$ rounds)
 connected graphs  $T^j=(V,F\cup F^j)$, $j\in[K]$, with the cost of
$O(\ln n +\ln K)L^*$. 
\begin{thm}
There is a randomized approximation algorithm for  \textsc{TSt-R Spanning Tree}
that yields  an $O(\log n +\log K)$-approximate solution, in general graphs, with high probability.
\label{tstrap}
\end{thm}
Therefore, when $K=\mathrm{poly}(n)$, Algorithm~\ref{alg2sstr} has the best approximation ratio
up to a constant factor
(see for an $O(\log n)$  (resp. $O(\log K)$) lower bound given~\cite{KZ11}).

In order to obtain a randomized $O(\log n +\log K)$-approximation algorithm for \textsc{TSt-E Spanning Tree}
it  is sufficient to solve in Step~\ref{slp} of Algorithm~\ref{alg2sstr}  a linear programming 
$\mathcal{LP}(L^*)$ counterpart  for   \textsc{TSt-E Spanning Tree}.
The analysis of such modified algorithm is similar to the previous one.
Now the total cost of  $\{T^j\}_{j\in [K]}$ returned  is as follows.
 \begin{lem}
 Let $\psi$ be the event that
  $\sum_{i\in F} C_i+\sum_{j\in [K]}\sum_{i\in F^j}p_jc_{ij}>\hat{k}L^*+(\mathrm{e}-1)\sqrt{\hat{k}L^{*}\ln(2Kn^2)}$,
  where $\hat{k}=\lceil   40\ln n +16\ln K \rceil$. 
  Then $\mathrm{Pr}[\psi]<\frac{1}{2Kn^2}$.
   \label{lebou1}
\end{lem}
Lemmas~\ref{lebou1}  and~\ref{lbou2}  and the union bound  show that Algorithm~\ref{alg2sstr} modified 
returns connected graphs  $T^j=(V,F\cup F^j)$, $j\in[K]$, with the cost of
$O(\ln n +\ln K)L^*$  with probability at least $1-\frac{1}{Kn^2}$.
\begin{thm}
There is a randomized approximation algorithm for  \textsc{TSt-E Spanning Tree}
that yields  an $O(\log n +\log K)$-approximate solution, in general graphs, with high probability.
\label{tsteap}
\end{thm}
In this case we also obtain the best approximation ratio
up to a constant factor
(see for an $O(\log n)$  (resp. $O(\log K)$) lower bound given~\cite{FFK06}).
From Theorem~\ref{thmgen0} it may be concluded that
\begin{cor}
There is a randomized $O(\log n +\log K)$-approximation algorithm for  
$\textsc{TSt-CVaR}_{\alpha}$ \textsc{Spanning Tree}
 for any constant $\alpha\in [0,1)$.
 \label{tstcvap}
\end{cor}

\subsection{Minimum assignment}

Let $G=(V,E)$ be a bipartite graph such that $V=A \cup B$ and $i\in A$, $j\in B$ for each $\{i,j\}\in E$. Let $\mathcal{X}$ be the set of characteristic vectors of the perfect matchings (assignments) in $G$. The problems $\textsc{TSt-E Assignment}$ and \textsc{TSt-R Assignment} for set $\overline{\mathcal{X}}$ were investigated in~\cite{KMU08}, where it was shown that $\textsc{TSt-E Assignment}$ is NP-hard for $K=2$ and hard to approximate within $O(\log n)$. If $|A|=|B|=n$ then \textsc{TSt-E Assignment} is approximable within $n^2$. Also, \textsc{TSt-E} and \textsc{TSt-R Assignment} are approximable within $1/\beta$ if a fraction of at least $n(1-\beta)$ nodes are matched. These approximation results are valid for set $\overline{\mathcal{X}}$, i.e. we only require that the set of edges, chosen in the first and second stage, contains an assignment.  In this section we will strengthen the results obtained in~\cite{KMU08} and provide new ones.

\begin{thm}
\label{thmass0}
	\textsc{TSt-R~Assignment} is NP-hard for $K=2$ and strongly NP-hard for unbounded $K$, even if $|\mathcal{X}|=|\overline{\mathcal{X}}|=1$, i.e. if there is only one feasible solution.
\end{thm}
\begin{proof}
	The idea is similar to the proof of Theorem~\ref{thmsp1}. It is enough to observe that the instance of \textsc{TSt-R~Assignment}, for the network shown in Figure~\ref{fig5}, is equivalent to \textsc{TSt-R~RS}.
	
	\begin{figure}[ht]
	\centering
	\includegraphics{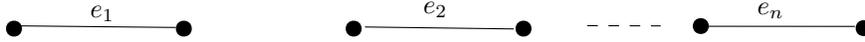}
	\caption{Illustration of the proof of Theorem~\ref{thmass0}.} \label{fig5} 
	\end{figure}
\end{proof}

\begin{lem}
\label{thmass2}
	There is a cost preserving reduction from \textsc{TSt-E(R) Shortest Path} with set $\mathcal{X}$ ($\overline{\mathcal{X}})$ and $K$ scenarios to \textsc{TSt-E(R) Assignment}  with set $\mathcal{X}$ ($\overline{\mathcal{X}})$ and $K$ scenarios.
\end{lem}
\begin{proof}
	We will use a well know one-to-one correspondence between simple $s-t$ paths in a network $G=(V,A)$ and assignments in a bipartite graph $G'=(V',E)$ (see, e.g.~\cite{AMO93}). Let $G=(V,A)$ be a network with $s,t\in V$, first stage arc costs $C_a$, $a\in A$, and the second stage arc costs $c_{aj}$ of $a\in A$, under scenario $j\in [K]$. We build the corresponding bipartite graph $G'=(V',E)$ as follows. The set of nodes $V'$ contains nodes $s$, $t'$ and $i$, $i'$ for each $i\in V\setminus\{s,t\}$. If $a=(i,j)\in A$, then we create the edge $\{i, j'\}\in E$. For each node $i\in V\setminus\{s,t\}$, we add \emph{dummy edge} $\{i,i'\}\in E$. An example is shown in Figure~\ref{fig10}.

	\begin{figure}[ht]
	\centering
	\includegraphics{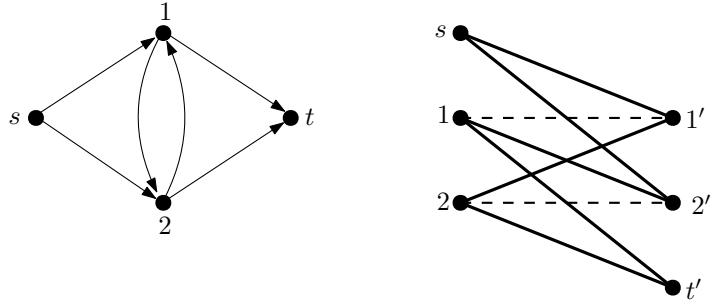}
	\caption{Illustration of the proof of Theorem~\ref{thmass2}. The dummy (deashed) edges have first stage costs equal to $M$ and the second stage costs equal to~0 under all scenarios.} \label{fig10} 
	\end{figure}
	
	There is a one-to-one correspondence between simple $s-t$ paths in $G$ and the assignments in $G'$. 	
	Namely, the arcs $(s,i_1), (i_1,i_2)\dots, (i_l, t)$ form a simple path in $G$, if and only if the edges $\{s,i_1'\}, \{i_1, i_2'\},\dots, \{i_l, t\}$ plus $\{i,i'\}$ for each $i\in V\setminus\{s,t,i_1,\dots,i_l\}$  form a perfect matching in $G'$. The first and the second stage costs of edge $\{i,j\}\in E$ are the same as the first and the second stage costs of the corresponding arc $(i,j)\in A$.
	The first stage costs of the dummy edges are set to $M=|A|\cdot c_{\max}$, where $c_{\max}$ is the maximum cost that appears in  the problem instance  (so no dummy edge is chosen in the first stage) and the second stage costs of these edges under each scenario are equal to~0 (so they do not influence on the cost of the recourse action).  The probability distribution in the resulting scenario set is the same is in the former one. Observe that there is one-to-one correspondence between the arcs in $G$ and not dummy edges of $G'$. Namely, $(i,j)\in A$ corresponds to $\{i,j'\}\in E$. Because no dummy edge can be chosen in the first stage and the costs of the dummy edges under all scenarios are 0, the lemma follows.
	
		\end{proof}

Using Lemma~\ref{thmass2} and the results obtained in Section~\ref{secsp} we get the following results:
\begin{thm}
\label{thmass1}
	\textsc{TSt-E} and \textsc{TS-R Assignment} with set $\mathcal{X}$ are strongly NP-hard and not at all approximable for $K=2$.
\end{thm}
\begin{cor}
	$\textsc{TSt-CVaR}_{\alpha}~\textsc{Assignment}$ with set $\mathcal{X}$ is strongly NP-hard and not at all approximable for $K=2$ and any fixed $\alpha\in [0,1)$.
\end{cor}

\begin{thm}
\label{thmass3}
	\textsc{TSt-E} and \textsc{TSt-R Assignment} with set $\overline{\mathcal{X}}$ are strongly NP-hard for $K=2$. Furthermore, for unbounded $K$ the problems are hard to approximate within $O(\log K)$.
\end{thm}

\begin{cor}
	$\textsc{TSt-CVaR}_{\alpha}~\textsc{Assignment}$ with set $\overline{\mathcal{X}}$ is strongly NP-hard for $K=2$ and any fixed $\alpha\in [0,1)$. Furthermore, for unbounded $K$ the problem is also hard to approximate within $O(\log K)$.
\end{cor}


\section{Conclusions}
\label{secconclusions}

In this paper we have investigated the complexity of two-stage versions of basic combinatorial optimization problems. It turns out that the two-stage approach leads to complex problems even in very restrictive cases. The negative results obtained can be used to characterize the computational properties of more complex problems. In particular, the selection problems are often special cases of more general models.
 A summary of the known and new results is shown in Table~\ref{tab2}.
\begin{table}[ht]
\begin{center}
\footnotesize
\caption{Summary of the known and new results. The symbols P means polynomially solvable,  and $\rho=\min\{\frac{1}{{\rm Pr}_{\min}},\frac{1}{1-\alpha}\}$. The results for the assignment problem are the same as for the shortest path.} \label{tab2}
\begin{tabular}{rr|ll}
\textsc{TSt} Problem & Criterion & $K\geq 2$ constant & $K$ unbounded \\
\hline
\textsc{RS} &\textsc{E} & P & P  \\[1.5ex]
& \textsc{R} & NP-hard & str. NP-hard, 2-appr.\\[1.5ex]
& $\textsc{CVaR}_{\alpha}$ & NP-hard for $\alpha\in [0.5,1)$ & $\min\{2,\frac{1}{1-\alpha}\}$-appr.\\
&					   & ? for $\alpha\in (0,0.5)$ \\
\hline
\textsc{Selection} & \textsc{E}& P  & str. NP-hard,  \\
&		   &					& not $O(\log n)$-appr.\\
&               &                                    &$O(\log n + \log K)$-appr.
\\[1.5ex]
& \textsc{R} & NP-hard \cite{KZ15b} & str. NP-hard~\cite{KZ15b},\\
 & & & not $O(\log n)$-appr. \cite{KZ15b}\\
 & & & $O(\log n + \log K)$-appr.~\cite{KZ15b}\\[1.5ex]
& $\textsc{CVaR}_{\alpha}$ & $\rho$-appr. & str. NP-hard for $\alpha\in [0,1)$ \\
& & & not $O(\log n)$-appr. for $\alpha\in [0,1)$\\
&&&   $O(\log n + \log K)$-appr. for $\alpha\in [0,1)$
\\
\hline
\textsc{Short. Path} ($\mathcal{X}$) & \textsc{E} & str. NP-hard, not appr. & \\
	&						   & P in sp-graphs & P in sp-graphs \\[1.5ex]
& \textsc{R} & str. NP-hard, not appr. & \\[1.5ex]
& $\textsc{CVaR}_{\alpha}$ & str. NP-hard, not appr. for $\alpha\in [0,1)$ & \\
&					   & $\rho$-appr. in sp-graphs & $\rho$-appr. in sp-graphs \\
\hline
\textsc{Short. Path} ($\overline{\mathcal{X}}$) & \textsc{E} & str. NP-hard & not $O(\log K)$-appr.\\[1.5ex]
& \textsc{R} & str. NP-hard & not $O(\log K)$-appr. \\[1.5ex]
& $\textsc{CVaR}_{\alpha}$ & str. NP-hard for $\alpha\in [0,1)$ & not $O(\log K)$-appr. for $\alpha \in [0,1)$\\
\hline
\textsc{Span. Tree} & \textsc{E} & ? & str. NP-hard~\cite{FFK06}, \\
& & & not $O(\log n)$-appr.~\cite{FFK06} \\
& & & $O(\log n+\log K)$-appr.\\[1.5ex]
& \textsc{R} & NP-hard  & str. NP-hard~\cite{KZ11} \\
&		  &	             & not $O(\log n)$-appr.~\cite{KZ11} \\
&		  &	              & $O(\log n+ \log K)$-appr. \\[1.5ex]
& $\textsc{CVaR}_{\alpha}$ & NP-hard for $\alpha\in [0.5,1)$ & str. NP-hard for $\alpha\in [0,1)$ \\
&					   & ? for $\alpha\in [0,0.5)$ & not $O(\log n)$-appr. for $\alpha \in [0,1)$ \\
& 					   &        				  & $O(\log n+ \log K)$-appr. for $\alpha\in [0,1)$
\end{tabular}
\end{center}
\end{table}

There is still a number of open questions concerning the considered problems. The complexity of $\textsc{TSt-CVaR}_{\alpha}~\textsc{RS}$ for $\alpha\in (0, 0.5)$ is open. This problem can also admit an FPTAS for constant $K$.
The complexity of \textsc{TSt-E Spanning Tree} for constant $K$ is open. 
 Also, there is lack of positive approximation results for the shortest path problem with set $\overline{\mathcal{X}}$. For the shortest path problem with set $\mathcal{X}$, the strong negative result hold for general graphs. Notice that for a special class of series-parallel graphs, the problem is easier.  So, it is interesting to explore the approximability of the problem for some special classes of graphs, for example for acyclic graphs. For unbounded $K$ randomized approximation algorithms for particular problems are known. It is interesting to check the existence of deterministic approximation algorithms for this case.

\subsubsection*{Acknowledgements}
The second and third author were supported by
 the National Science Centre, Poland, grant 2017/25/B/ST6/00486.


\appendix

\section{Some proofs}
\label{dod}

\begin{proof}[Proof of Lemma~\ref{lsalon}]
If  $P^j_{k-1}=0$, then 
we are done.  Assume that $P^j_{k-1}\geq 1$ and
consider the set of items~$E^j_{k-1}$, $|E^j_{k-1}|=N^j_{k-1}$ and
the number of items $P^j_{k-1}$, remaining for selection in round $k$.
Let $\mathrm{I}_i$ be a random variable such  
that $\mathrm{I}_i=1$ if item~$i$ is  picked from $E_{k-1}^j$; 
and $\mathrm{I}_i=0$, otherwise. It is easily seen that
$\mathrm{Pr} [\mathrm{I}_i=1]=1-(1-x^*_i) (1-y^*_{ij})$.
The expected number of items selected out of~$E^j_{k-1}$ in round $k$
is
\begin{align*}
\mathbf{E}\left[\sum_{i\in E^j_{k-1}} \mathrm{I}_i\right]&=
\sum_{i\in E^j_{k-1}}\mathrm{Pr}\left[\mathrm{I}_i=1\right]=
N^j_{k-1}-\sum_{i\in E^j_{k-1}}(1-x^*_i) (1-y^*_{ij})\\
&\geq 
N^j_{k-1}-\sum_{i\in E^j_{k-1}}\left(1-\frac{x^*_i+y^*_{ij}}{2} \right)\geq
\frac{P^j_{k-1}}{2}.
\end{align*}
The first inequality follows from the fact that $ab\leq(a+b)/2$ for any $a,b\in [0,1]$.
The second one follows from the fact that  the feasible  solution~$(\pmb{x}^*,\pmb{y}^*)$
satisfies  constraints~(\ref{hcs}).
Using  Chernoff bound (see, e.g.,\cite[Theorem~4.5  for
$\delta=\sqrt{4\ln 1.25/P^j_{k-1}}$]{MU05}), we get
\[
\mathrm{Pr}\left[\sum_{i\in E^j_{k-1}} \mathrm{I}_i < \frac{P^j_{k-1}}{2}-
P^j_{k-1}\sqrt{\frac{\ln 1.25}{P^j_{k-1}}}\right]\leq \frac{4}{5}.
\]
Thus, with  probability at 
least $1/5$, the number of selected items in round $k$  is at least
$P^j_{k-1}/2-\sqrt{P^j_{k-1}\ln 1.25}$.
 Hence, with probability at least $1/5$ it holds
\[
P^j_k\leq P^j_{k-1}-P^j_{k-1}/2+\sqrt{P^j_{k-1}\ln 1.25}=(1/2+\sqrt{\ln 1.25/P^j_{k-1}})P^j_{k-1}.
\]
Consequently, when $P^j_{k-1}\geq 1$ we get
$P^j_k<0.98 P^j_{k-1}$
with probability at least $1/5$.
\end{proof}

\begin{proof}[Proof of Lemma~\ref{lsbou2}]
Let us estimate the number $\ell$ of successful rounds, among $\hat{k}$~performed rounds,
which are sufficient to ensure
$P^j_{\hat{k}}=0$.  We must have $(0.98)^{\ell} p <1$.
The above inequality holds, in particular, when $\ell > 50\ln p$.
Let $\mathrm{Z}_k$ be a binary random variable such that
$\mathrm{Z}_k=1$ if and only if round~$k$ is  ``successful'', $k\in [\hat{k}]$.
 In order to cope with the dependency of the events: 
round~$k$ is  ``successful'',
$\mathrm{Pr}[\sum_{k\in [\hat{k}]}\mathrm{Z}_k\leq50\ln p]$ is estimated  from above by 
$\mathrm{Pr}[\mathrm{B}(\hat{k},1/5)\leq 50\ln n]$, where
$\mathrm{B}(\hat{k},1/5)$ is a binomial random variable (see
Lemma~\ref{lsalon} and~\cite[Lemma~14.6]{MU05}). Thus
applying  Chernoff bound 
(see, e.g.,
\cite[Theorem~4.5]{MU05} and
$\hat{k}= \lceil 335\ln n+ 40\ln 2K \rceil$) and the  fact that $p<n$
 yield the following upper bound on $\mathrm{Pr}[\xi^j]$:
\begin{align*}
\mathrm{Pr}[\xi^j]&\leq \mathrm{Pr}\left[\sum_{k\in [\hat{k}]}\mathrm{Z}_k\leq 50\ln p\right]\leq
\mathrm{Pr}[\mathrm{B}(\hat{k},1/2)\leq 50\ln p]
< \mathrm{e}^{-(17\ln n+8\ln 2K)^2/(2(67\ln n +8\ln 2K))}\\
&\leq\mathrm{e}^{-(17\ln n+8\ln 2K)/8}< \frac{1}{2Kn^2}.
\end{align*}
This proves the lemma.
\end{proof}

\end{document}